\documentclass[11pt]{article}
\usepackage{pifont}
\usepackage{natbib}
\usepackage{graphicx}
\usepackage{amssymb}
\usepackage{epstopdf}
\usepackage{amsmath}
\usepackage{subfigure}
\usepackage{lscape}
\usepackage{latexsym}
\usepackage{amsmath}
\usepackage{yhmath}
\usepackage{bm}
\usepackage{epsfig}
\usepackage{fancyhdr}
\usepackage{sectsty,lscape,url}
\usepackage[hmargin=1in,vmargin=1in]{geometry}
\usepackage{graphicx}
\usepackage{sidecap}

\usepackage{pifont}
\usepackage{natbib}
\usepackage{graphicx}
\usepackage{amssymb}
\usepackage{epstopdf}
\usepackage{amsmath}
\usepackage{lineno}
\usepackage{setspace}
\usepackage{url}
\usepackage{lscape}
\usepackage{color}
\usepackage[normalem]{ulem}
\usepackage{sidecap}
\usepackage[font={small}]{caption}
\usepackage{authblk}

\begin{document}

\title{A response function approach for rapid far-field tsunami forecasting}

\author[1]{Elena Tolkova \thanks{e.tolkova@gmail.com; elena@nwra.com}}
\author[2]{Dmitry Nicolsky \thanks{djnicolsky@alaska.edu}}
\author[3]{Dailin Wang \thanks{dailin.wang@noaa.gov}}
\affil[1]{NorthWest Research Associates, Bellevue, WA 98009-3027, USA}
\affil[2]{University of Alaska Fairbanks, Fairbanks, AK 99709, USA}
\affil[3]{NOAA/NWS/Pacific Tsunami Warning Center, Honolulu, HI, 96818, USA}
\date{ }
\maketitle

{\emph{The final publication is available at Springer via}} {\url{https://link.springer.com/article/10.1007/s00024-017-1612-0}}
\tableofcontents
\bigskip
\begin{abstract}
Predicting tsunami impacts at remote coasts largely relies on tsunami en-route measurements in an open ocean.
In this work, these measurements are used to generate instant tsunami predictions in deep water and near the coast. 
The predictions are generated as a response or a combination of responses to one or more tsunameters, with each response obtained as a convolution of real-time tsunameter measurements and a pre-computed Pulse Response Function (PRF).
Practical implementation of this method requires tables of PRFs in a 3D parameter space: earthquake location - tsunameter -  forecasted site. Examples of hindcasting the 2010 Chilean and the 2011 Tohoku-Oki tsunamis along the US West Coast and beyond demonstrated high accuracy of the suggested technology in application to trans-Pacific seismically-generated tsunamis.
\end{abstract}


\section{Introduction}
\label{RnD}

Tsunami forecasting, in particular in application to ocean-wide seismically-generated tsunamis, relies on numerical modeling for predicting the tsunami wave height and currents before the actual tsunami arrives at the coasts. Conventionally, the modeling solves an initial value problem with an estimated tsunami source function. At the same time, deducing the tsunami source presents a major challenge. Existing instruments do not directly measure the tsunami source -- an initial displacement of the sea surface in the earthquake (EQ) area. Instead, the source has to be inferred from related data.

The first-available land-based seismic and geodetic observations can provide for rapid warning for nearby coasts within a few minutes after the EQ, with more elaborated source products to follow \cite[]{melgar2016}. 
The next available source estimate, produced by a technology termed W phase inversion of centroid moment tensor, uses measurements of long period (100-1000 s) seismic waves in the far field of the EQ. Since 2009, the Pacific Tsunami Warning Center (PTWC) routinely makes real-time tsunami forecasts taking the focal mechanism from the W phase inversion and assuming a uniform slip on a rectangular fault \citep{foster2012,wang2012}. Seismogeodesy with near-source real-time GPS measurements \citep{crowell2012, melgar2013, titov2016a} can improve characterization of the fault geometry, but its applicability is limited to EQs occurring in the vicinity of dense GPS arrays. This sequence of early to relatively early available source models is used for estimating expected tsunami heights by coarsely defined ranges assigned to coastal zones  (Melgar et al. (2016) and references therein). 

Once a tsunami reaches open-ocean tsunameter stations, sea level measurements offer an attractive option to bypass the complexity of fault motions, and to reconstruct the tsunami source directly. Commonly, the source is assembled from pre-defined basis functions. It is assumed that hypothetical tsunamis radiated by these basis sources linearly superimpose in deep water. Then the basis sources/tsunamis are combined to achieve the best fit to the observations - a procedure termed source inversion. The resulting source is expected to further advance the accuracy of the forecast, which can now aim at predicting location-specific time histories of wave elevation and currents. 

The latter approach is implemented within a tsunami forecasting system termed SIFT (Short-term Inundation Forecasting for Tsunamis), developed at PMEL (Pacific Marine Environmental Laboratory, NOAA) and operationalized at both US Tsunami Warning Centers.
The basis functions in SIFT are represented by ``unit" EQs \cite[]{gica, perc} with a fault length of 100 km, fault width of 50 km, and a pre-defined spatial orientation. This apparently coarse size of the basis functions and the prescribed EQ mechanism allow a user to obtain meaningful source inversions by fitting observations at as few as one or two deep-ocean stations. The observations are provided by the DART network comprising about 40 stations stretched along the Pacific Rim.
The SIFT technology resulted in a sequence of successful forecasts and hindcasts \citep{tang2008, tang2012, tang2016, wei2008}. 
At the same time, the inverted tsunami source is sensitive to the subset of considered ``unit" sources, choice of DARTs, and duration of the observations being fitted. As stated by \cite{perc}, ``proper selection of the unit sources and of the relevant data is vital for getting good results from the inversion algorithm" .

The SIFT methodology relies on studies by 
PMEL scientists who emphasize that ``only few source parameters are critical to characterize far-field tsunamis, mainly the location and the magnitude. Other parameters have secondary influence and can be predefined for forecast purposes" \citep{titov1999, tang2016}. 
In particular, ``the waves are not very sensitive to details of the initial ocean surface deformation, as long as the [wave] energy and the general source area are determined correctly" \cite[]{tang2012}. These findings imply that the method aims at constructing a source proxy, not necessarily the physical source, which would nevertheless approximate the tsunami waveform outside the near field.

Therefore, the next logical step is to by-pass the construction of the source proxy and use tsunami's en-route measurements for recovering its wavefront and project its dynamics forward in time. To date, a few such attempts have been made. These studies were directed at forecasting local coasts with the measurements collected in their vicinity, and provided warning times under 1-2 hours. 
Maeda et al. (2015) developed a data assimilation method to recover tsunami waves passing through a dense array of tsunameters. This methodology strongly focuses on using the Japan Trench cabled network with detectors spaced at 30-50 km.
Thomson et al. (2011) forecasted the 2009 Samoa tsunami at the coast of Vancouver Island, BC, Canada, using the wave measurements at the NEPTUNE cabled observatory off-shore the island as a realtime plane-wave input into a regional tsunami forecast model. 
Elaborating upon physical assumptions similar to that used by Thomson et al. (2011), Power and Tolkova (2013) expressed their forecast with a response formalism. The Pulse Response Function, as introduced by Power and Tolkova (2013), is unique for each ``detector-site" pair and for the directivity of the approaching tsunami, which is determined by the location of the tsunami source. The PRF does not depend on the details of the sea surface deformation in the tsunami origin. The response functions can be pre-computed for a set of anticipated approach directions corresponding to different tsunami origination areas, and then used in a tsunami event to generate instant real-time offshore and near-shore estimates.   

In this paper, we introduce a new tsunami forecasting method which generates a prediction at a far-field forecast site as a combination of convolutions of the tsunami's real-time measurements and pre-computed Pulse Response Functions (PRFs). As will be detailed further, the proposed method provides an instant solution to the boundary value problem of numerically propagating the wave directly from the detectors in a linear propagation medium. This technology relies on mathematical formalism developed in Power and Tolkova (2013), adapted for using measurements at any number of detectors in the relative vicinity of the source area, outside the near field. 

The new technology is described in section \ref{methods}, and is demonstrated with hindcasting the two largest trans-Pacific tsunamis of the 21st century: the 2010/02/27 Chile (section \ref{chile}) and the 2011/03/11 Tohoku-Oki (section \ref{tohoku}) tsunamis. 
In both cases, tsunami measurements at open ocean detector(s) in the EQ region are directly converted into the wave time histories at remote offshore locations across the ocean, where the wave would arrive many hours later. Next, the PRFs derived for the Tohoku-oki tsunami were ``blind-tested" with forecasting tsunamis generated by a hypothetical normal fault event and a hypothetical thrust event in the general area of the 2011 Tohoku-Oki EQ (section \ref{btest}). This exercise demonstrates that PRFs do not depend on the EQ mechanism and geometry, and are applicable to forecasting any tsunami originating in the same general area.
In section \ref{sgages}, the response-based forecast is extended to coastal sites. In particular, the response formalism is generalized for predicting wave heights in non-linear coastal environments such as a shallow harbor, and applied to hindcasting the above two tsunamis in the Crescent City Harbor, CA  (sections \ref{sgages1}-\ref{sgages3}). Forecast latency is  briefly discussed in section \ref{wtime}. 
Finally, Conclusions elaborate upon the potential of the suggested technology, its pros and cons compared to related conventional methods, and directions for further research.

\section{Mathematical and Numerical Methods}
\label{methods}

The response formalism relies on an assumption that the azimuthal (across-the-beam) shape of the tsunami front is prescribed by the bathymetry and a location of the tsunami source; whereas the radial (along-the-beam) wave structure is determined by the source details. Then a tsunami beam within a certain azimuthal window can be reconstructed by combining a prescribed wavefront shape with the radial wave structure measured as the wave passes by a detector. 
To reconstruct an azimuthal structure of the departing wave within a larger angle, or even ocean-wide, the wave has to be assembled from several wave beams. The latter requires several detectors delimiting the source region.

\subsection{Response formalism with a single detector}
\label{math}

\begin{SCfigure}
\centering
	\caption{Interpolating pulse $h(t)$, plotted against $t/\Delta t_0$. }
		\includegraphics[width=0.4\textwidth]{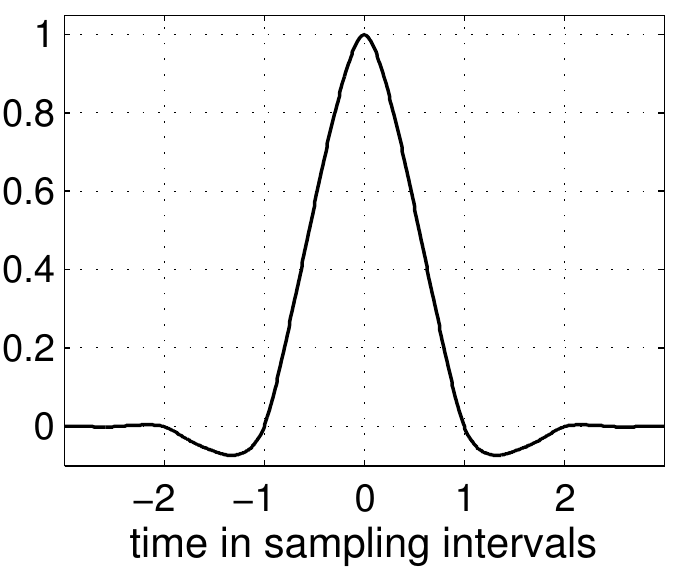}
	\label{plsh}
\end{SCfigure}

Power and Tolkova (2013) introduced the following formalism describing a wave beam evolution past a detector.
The measurements $q_i$ taken at a deep-ocean detector with an interval $\Delta t_0$ represent a discrete signal given by the series of delta-functions: 
\begin{equation}
q(t)=\sum{q_i \delta(t-i\Delta t_0)},
\label{dq}
\end{equation}
which can be converted into a continuous signal as 
\begin{equation}
\tilde{q}(t)= q(t) \otimes h(t) =\sum{q_i h(t-i\Delta t_0)}, 
\label{q}
\end{equation}
where $\otimes$ denotes convolution; $h(t)$ satisfies a series of conditions: $h(0)=1$ and $h(i\Delta t_0)=0$ for all integer $i \ne 0$; $\sum{ h(t-i\Delta t_0)} =1$; $\int {h(t)dt}=\Delta t_0$; and its spectrum is localized below $1/2\Delta t_0$. Impulse $h(t)$ acts over the series $q(t)$ as a low-pass filter with a cut-off at the Nyquist frequency. A function $h(t)$ with such properties is known as an interpolating pulse \cite[]{wvlets}; its particular example adopted for this work and shown in Figure \ref{plsh} is described in (Power and Tolkova, 2013). The sampling interval, selected hereafter at $\Delta t_0=2$ min, determines the period $2\Delta t_0$ of the shortest wave component registered at the detector. 

\begin{figure}[ht]
\begin{center}
	\resizebox{0.9\textwidth}{!} %
		{\includegraphics{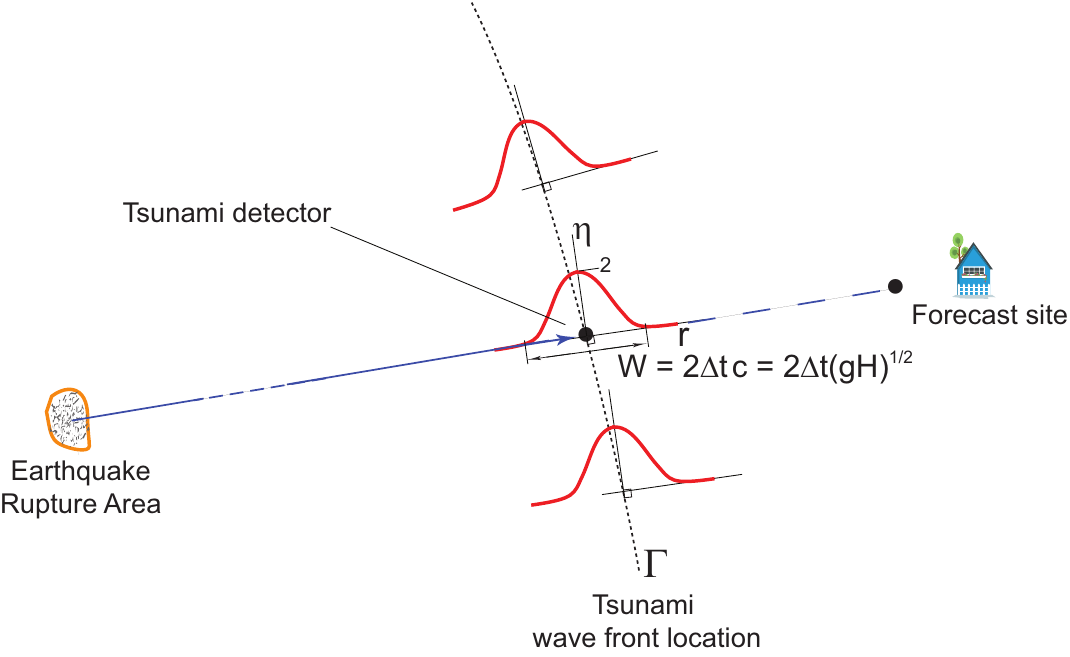}} 
	\caption{Structure of the initial deformation. The width of the initial shape between two zeros of the sea surface displacement, estimated as $W=2 \cdot c \cdot \Delta t_0$, is only 48 km at $H=4000$ m depth ($c=200$ m/s).}
	\label{pulse1}
	\end{center}
\end{figure}

Assuming linearity of the propagation, the tsunami signal (wave height or flow velocity) at a forecasted site is sought as 
\begin{equation}
s(t)=q(t) \otimes p(t) = \sum{q_i p( t-i\Delta t_0)}, 
\label{sq}
\end{equation} 
where $p(t)$ is the Pulse Response Function (PRF). It represents the tsunami signal at the site, divided by $h(0)$, in a hypothetical wave which propagates in the tsunami's direction and yields the surface elevation time history $h(t)$ at the detector (the interpolating pulse $h(t)$ in eq.\eqref{q} is dimensionless, but the surface elevation $h(t)$ is measured in meters. Hereafter, the dimensionality will be implied by the context and not shown explicitly).
In the half-space down the tsunami path from the detector, this hypothetical wave can be obtained as a solution to the SWE initialized in either of two ways:
\begin{enumerate}
\item
by varying the surface elevation as $2 \cdot h(t)$ with zero flow velocities on a boundary $\Gamma$ drawn through the detector and coinciding with the tsunami wave front; or 
\item
by a specific initial deformation centered on the detector. The deformation has an elongated shape containing the assumed boundary $\Gamma$ on its centerline, with the shape's profile across $\Gamma$ being $\eta=2 \cdot h(r/c)$, where $c$ is the long wave celerity offshore, and $r$ is a distance from the shape's centerline (Figure \ref{pulse1}). In this work, the same $c$ value is applied at all points on $\Gamma$.
\end{enumerate}
In a linear medium (deep water), either procedure produces two waves, each with the time history $h(t)$ at the detector. One wave travels toward the forecasted site, and the other -- back to the tsunami source.
If variations of the local wave celerity $c$ on $\Gamma$ are negligible, the above two options yield identical results. Then expression \eqref{sq} provides a solution at the site to an equivalent boundary value problem in an initially still domain, separated from the source area by curve $\Gamma$ and forced by an incoming wave with an elevation time history $\tilde{q}(t)$ on $\Gamma$.                                                 

In this work, PRF is computed as a wave-height time-history at a forecasted site obtained by solving an initial value problem (option 2 above) for the fully-nonlinear SWE. It is worth noting that:
\begin{itemize}
\item
The solution to the above problem should also yield a time-history $2 \cdot h(t)$, $t \ge 0$, at the detector -- the result of superposition of the left half on the initial disturbance propagating to the right and the right half propagating to the left. 
Then an offshore wave celerity $c$ might need to be adjusted (and the initial condition re-derived), to ensure that the first zero surface displacement at the detector occurs at exactly $\Delta t_0$. The latter condition defines the value of $c$ regardless of its initial estimate (for instance, with the depth at the detector).
\item
Half of the initial disturbance tends to focus back on the source. For the PRF computations, the bathymetry/topography in the source area might need to be modified, to allow the back-traveling pulse to exit the domain without being reflected onto the detector (particular bathymetry modifications used in this study are described in the Supplementary Appendix). 
\item
The implication of linearity of the propagation holds when non-linear effects in the SWE solutions are negligible, which is justified when the forecasted site is located in deep water or the tsunami height remains small.
\item
An actual tsunami wave is unlikely to be uniform along its entire wavefront. Therefore expression \eqref{sq} can be physically meaningful only within a certain azimuthal window enclosing the detector, and the resulting prediction applies only to sites located in the path of the wave beam within this window.
\end{itemize}

\subsection{Response formalism with multiple detectors: combining wave beams} 
\label{Nmath}

\begin{SCfigure}
\centering
	\caption{Detectors (triangles) provide boundary input into the domain on the right, separated from the source area by thick (solid or dashed) lines, composed of segments of rays (dashed) and wavefronts (solid) emanating from a point source in the EQ epicenter. An arbitrary point $P$ is defined by its  distances $r$ and $\rho$ from the $k$-th detector across and along the wavefront $\Gamma_k$. }
		\includegraphics[width=0.5\textwidth]{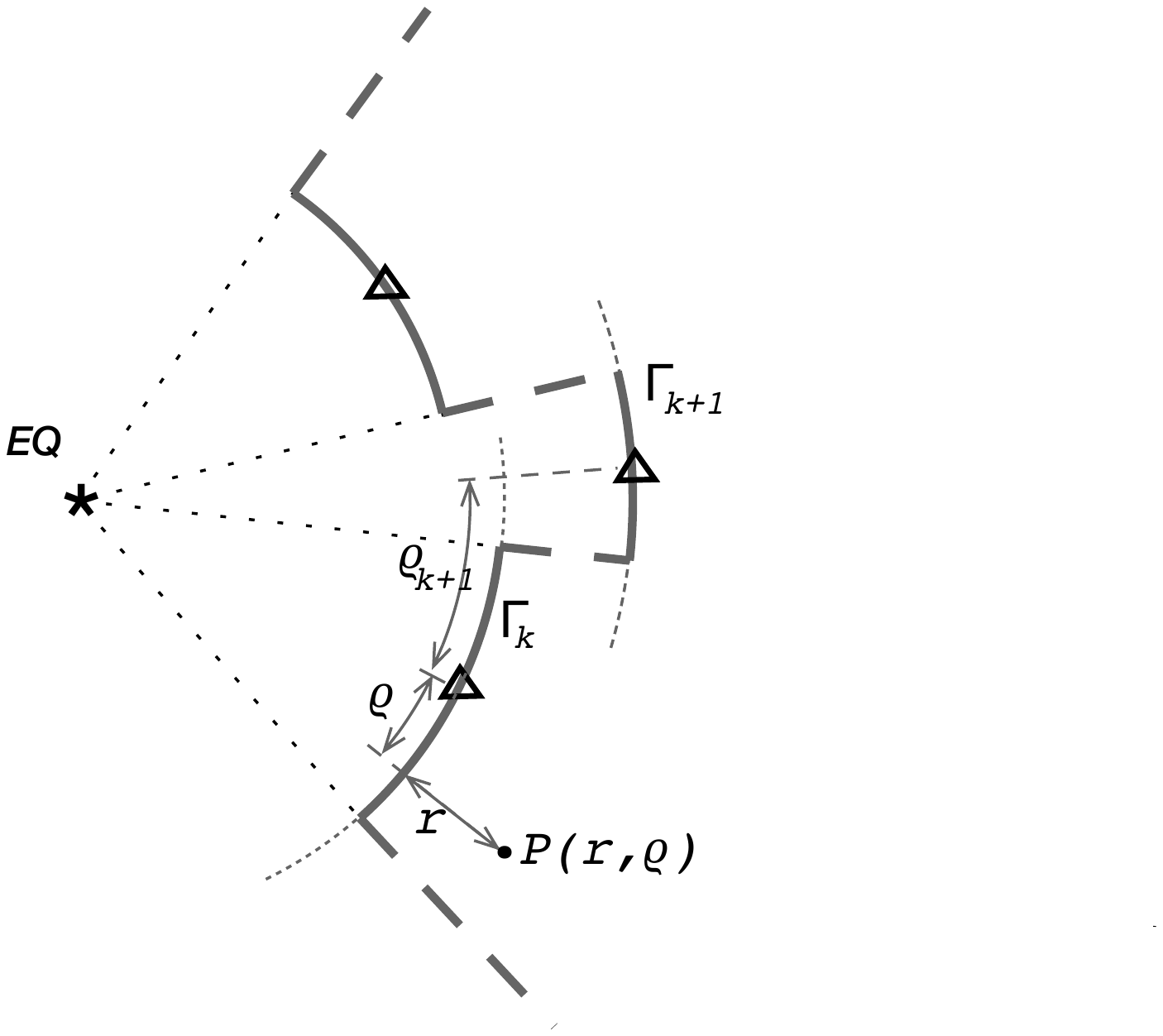}
	\label{diag}
\end{SCfigure} 

Several detectors delimit the source region with an imaginary boundary composed of wavefront segments connected with wave rays, as depicted in Figure \ref{diag}.
No wave energy propagates across the rays. Therefore, under the assumption of linearity of the propagation tract, the signal at the site is a superposition of responses $s_k$, $k=1, \dots, K$, to the wave forcing through the disconnected wavefront segments $\Gamma_k$ associated with $K$ individual detectors: 
\begin{equation}
s(t)=\sum_{k=1}^{K} {s_k(t) }, 
\label{ss}
\end{equation} 
where each detector's contribution $s_k$ is a convolution \eqref{sq} of this detector's measurements and its PRF. The PRF for each detector can be computed as a solution to an initial value problem, with the initial deformation built upon the $k$-th wavefront segment, as described in section \ref{math}.

Furthermore, in assembling the wave from $K$ wave beams and making it continuous in the azimuthal direction, one might want to make the beams overlapping and tapered across with a function $\tilde{h}(\rho)$, where $\rho$ is a distance along $\Gamma_k$.
As a result, an initial deformation $\eta_0(P)$ at point $P$ for the $k$-th PRF computations is given by:
\begin{equation}
\eta_0(P) = 2 \cdot h(r/c) \cdot \tilde{h}_k(\rho),
\label{shape}
\end{equation}
where $r$ and $\rho$ are distances of the point $P$ from the $k$-th detector counted across and along the wavefront $\Gamma_k$, as shown in Figures \ref{pulse1} and \ref{diag}; other notations are the same as in section \ref{math}.
Azimuthal interpolator $\tilde{h}_k(\rho)$ describes a spatial span of the $k$-th wavefront segment and an optional azimuthal tapering factor at the $k$-th detector. Anticipated spacial spans can vary from non-overlapping segments (as shown in Figure \ref{diag}) to wider segments delimited at the $k$-th detector by the rays propagating to the $(k+1)$-th and $(k-1)$-th detectors. The latter option was adopted in this work. If no wavefront singularity occurs between the $k$-th and $(k+1)$-th detectors, the linear interpolation is applied: $\tilde{h}_k(\rho)=1-\rho/\rho_{k+1}$, $0<\rho<\rho_{k+1}$, where $\rho_{k+1}$ is a distance at which $(k+1)$-th ray passes by the $k$-th detector. If both $k$-th and $(k+1)$ detectors are located in deep water with no local bathymetric peculiarities, $\rho_{k+1}$ can be approximated by the distance of the $(k+1)$-th detector's projection onto the $k$-th wavefront (see Figure \ref{diag}). With $\tilde{h}_{k+1}(\rho)$ selected analogously, the $k$-th and the $(k+1)$-th wave beams continuously transition into each other. 
On the contrary, if there is a ridge or another topographic feature breaking a wavefront at a distance $\rho_{ridge}$ from the $k$-th detector, the initial deformation extends uniformly to the ridge and ends there: $\tilde{h}_k (\rho<\rho_{ridge})=1$, $\tilde{h}_k (\rho>\rho_{ridge})=0$. 

\subsection{Numerical Models}

Three numerical models solving the Shallow-Water Equations (SWE) on a sphere are used in this study for PRF computations (Cliffs, Alaska GI-T) and for forecasting with simulated events (RIFT).

The Cliffs model is an open-source relative of the MOST (Method of Splitting Tsunamis) model used in the SIFT system. As MOST, Cliffs solves the fully nonlinear SWE in a characteristics form with an explicit finite-difference scheme developed at the Novosibirsk Computing Center of the Siberian Division of the Russian Academy of Sciences in 1984-1989 \cite[]{titov-most2016}. The friction term is based on Manning formulation. The primary difference between Cliffs and MOST algorithms is the treatment of the land-water boundary \citep{cliffs, cliffsbp}.

The Alaska GI-T model stems from the TUNAMI model by Imamura (1996), which solves the fully nonlinear SWE with a leap-frog numerical scheme; friction in Manning formulation. The primary modification in the Alaska GI-T model is the method used for wetting/drying \cite[]{nicolsky}. Currently, the Alaska GI-T model is used to predict propagation and runup of hypothetical tsunamis along the Alaska shore \citep{suleimani2013, nicolsky2015}.

The RIFT model is a real-time tsunami forecast model, developed in the US Pacific Tsunami Warning Center (PTWC) for rapid ocean-scale simulations \cite[]{wang2012}. The model solves the linearized SWE with a finite difference scheme similar to one used in Imamura (1996), without friction.

\section{Case study: Forecasting the 2010 Chile tsunami along North America from a DART off the coast of Peru}
\label{chile}

\begin{figure}[ht]
\begin{center}
	\resizebox{\textwidth}{!} %
		{\includegraphics{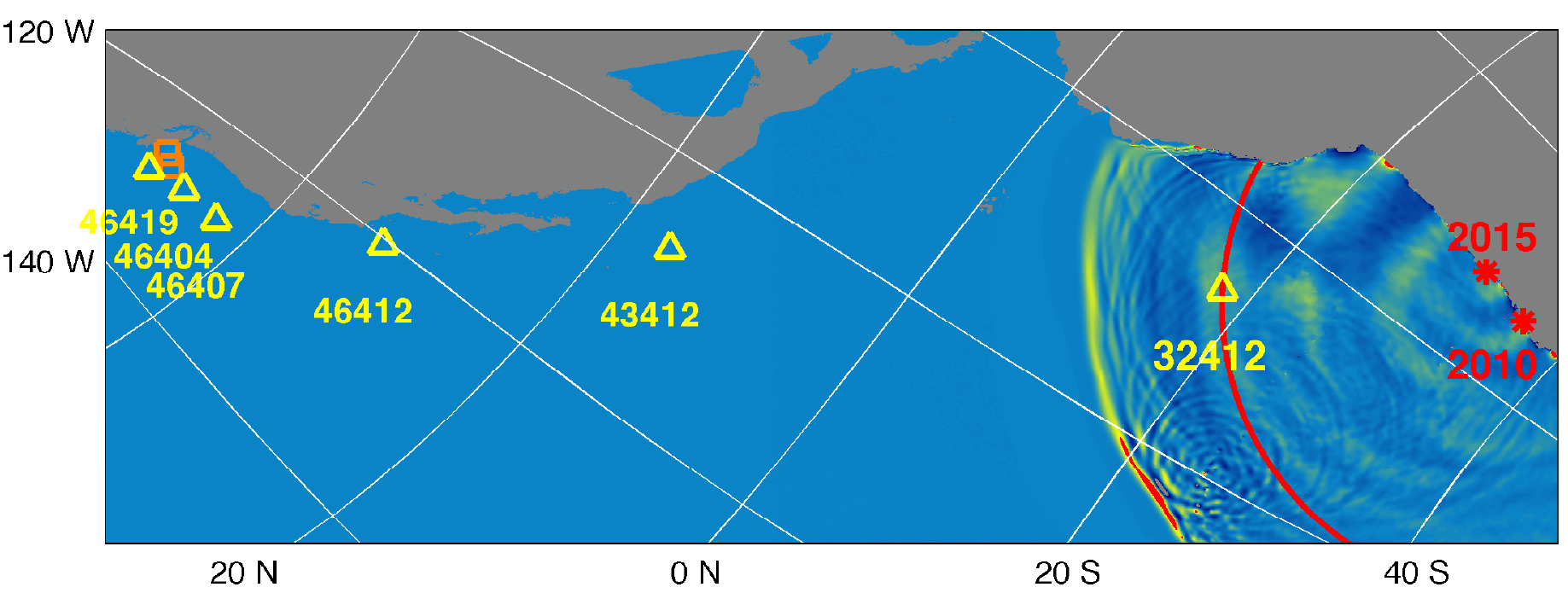}} 
	\caption{Sea surface snapshot from a simulation of the 2010/02/27 tsunami. Triangles: DART station 32412 offshore Peru; five West Coast DARTs (south to north, or right to left in the map) 43412 (Mexico), 46412 (CA), 46407 (OR), 46404 (OR-WA), and 46419 (WA); squares: two ONC BPRs 22503 at 860 m depth and CORK at 2660 m depth (BC, Canada). Red arc: wavefront estimate $\Gamma$. Stars: epicenters of the 2010/02/27 and 2015/09/16 EQs.}
	\label{mapCh}
	\end{center}
\end{figure}

Northern and Central Chile is known for its tsunamigenic EQs, including the Mw 9.5 EQ of May 22, 1960 -- the strongest instrumentally recorded EQ in history, and the more recent EQs of $M_w$ 8.8 on 2010/02/27, $M_w$ 8.2 on 2014/04/01, and $M_w$ 8.3 on 2015/09/16. 
The north-bound beam of each Chilean tsunami travels to the North American coastline, recorded by DARTs along the way. Below, measurements of the 2010/02/27 Chile tsunami at DART station 32412 offshore Peru are converted into the wave time histories along the coast of North America from Mexico to Canada (Figure \ref{mapCh}), 6.5-13 hours before the tsunami would arrive there. The single-detector version of the response method (section \ref{math}) is sufficient here, because all waves carrying the bulk of the tsunami energy north, including the later waves, fall within the same wave beam encompassing the DART station; whereas no significant waves from Chile can arrive on the US West Coast following other routes.

\begin{figure}[ht]
\begin{center}
	\resizebox{0.7\textwidth}{!} %
		{\includegraphics{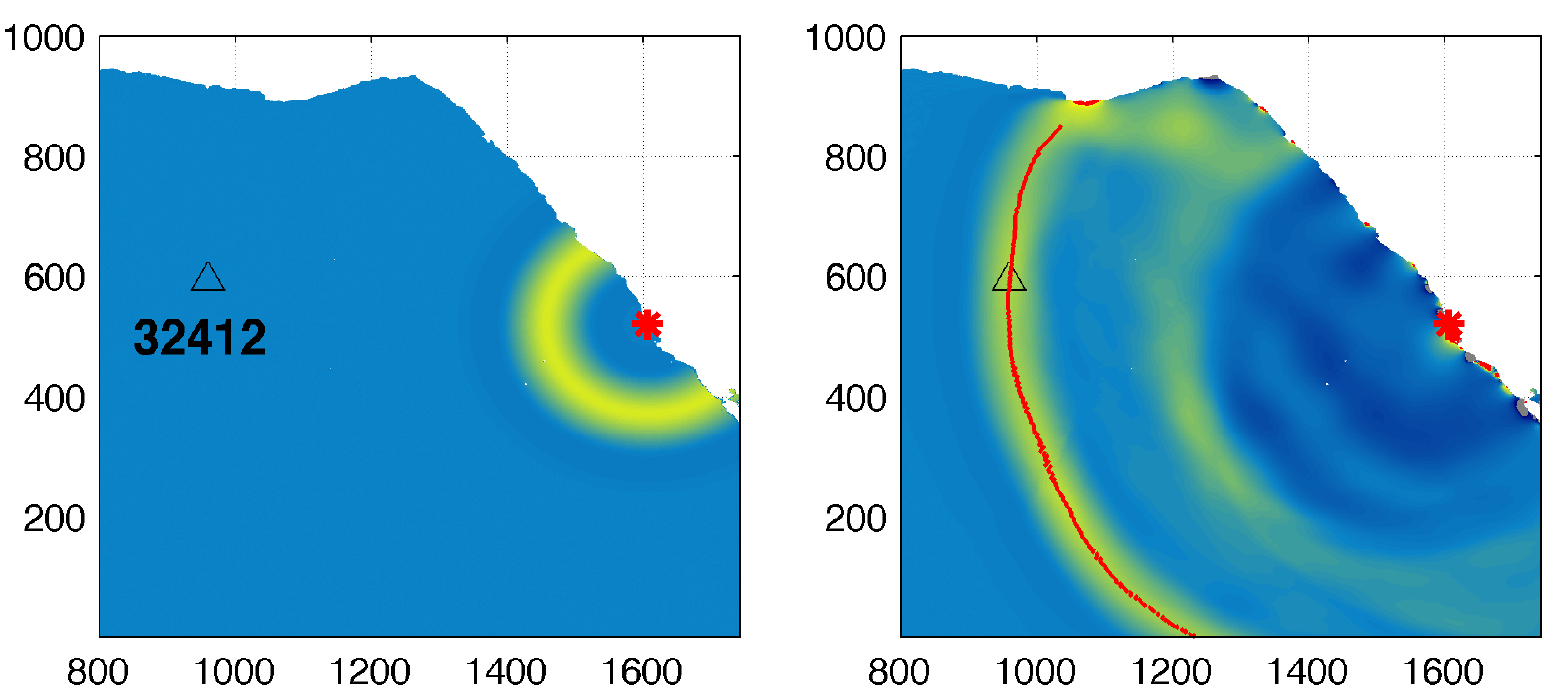}} 
	\caption{Left: initial condition to compute wavefront $\Gamma$. Right: snapshot from the simulation when the wave crest reached 32412; points under the wave crest (red line) define curve $\Gamma$.
Star: epicenter of the 2010/02/27 EQ. Map projection as in Figure \ref{mapCh}. Axis: node numbers, on a grid with a 2 arc-min spacing.}
	\label{mapgamma}
	\end{center}
\end{figure} 

\begin{figure}[ht]
\begin{center}
\begin{tabular}{cc}
	\includegraphics[width=0.98\textwidth]{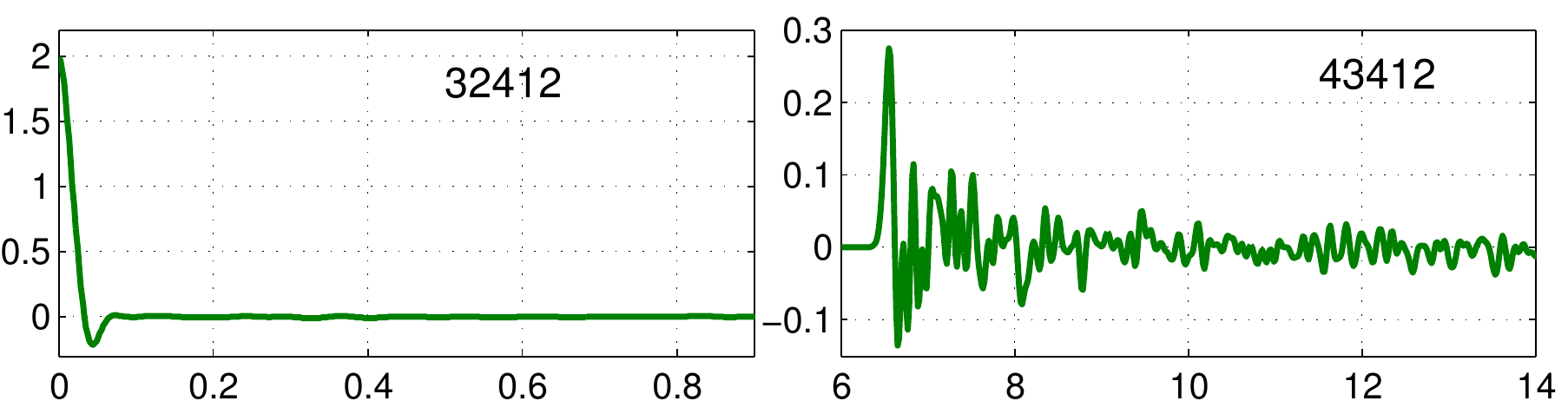} \\
	\includegraphics[width=0.49\textwidth]{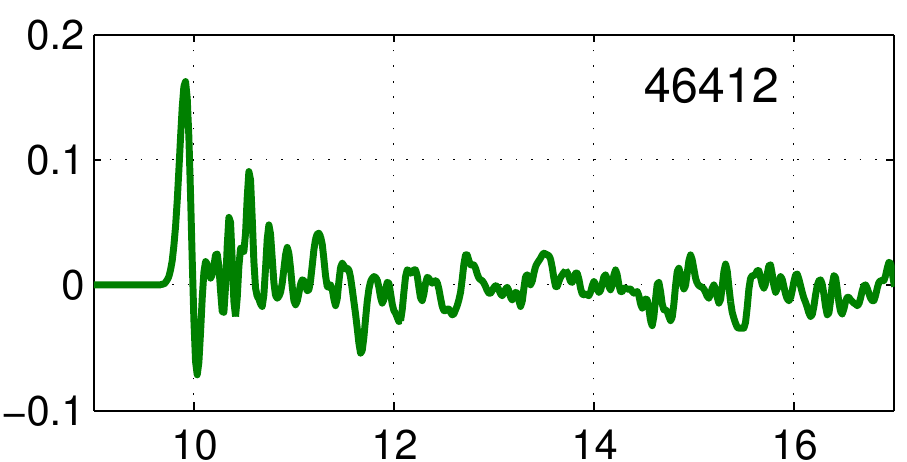} 
	\includegraphics[width=0.49\textwidth]{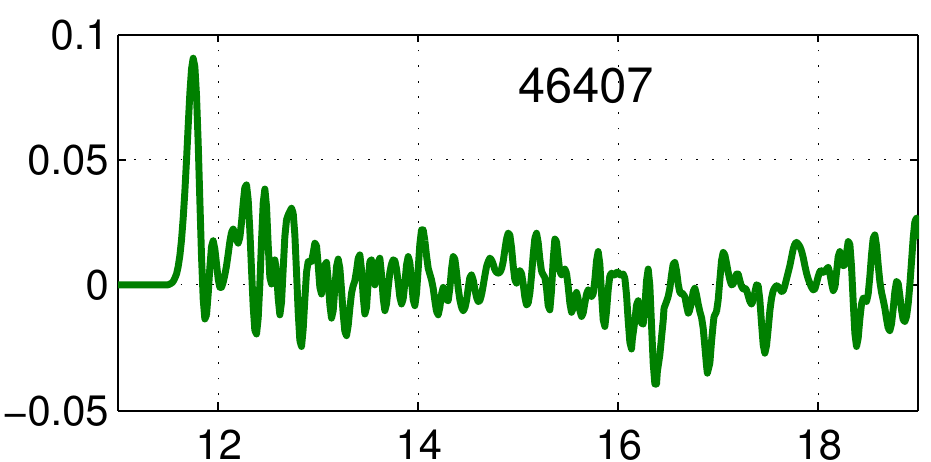} \\
	\includegraphics[width=0.5\textwidth]{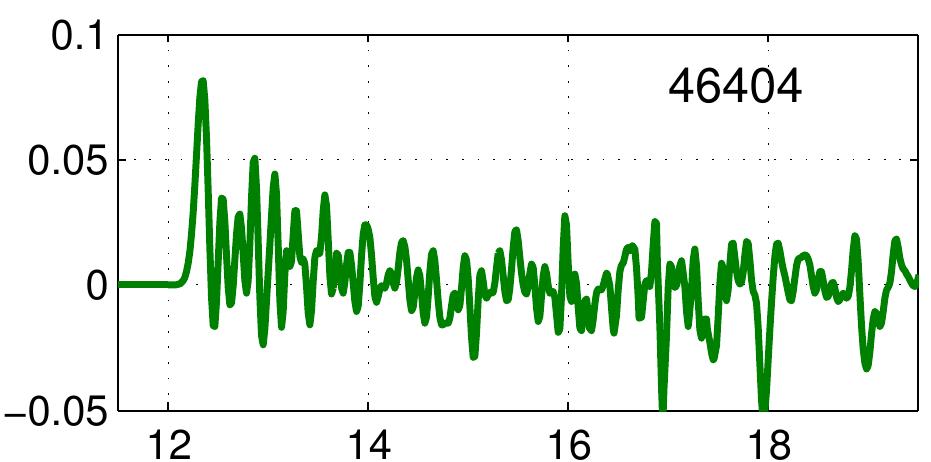} 
	\includegraphics[width=0.5\textwidth]{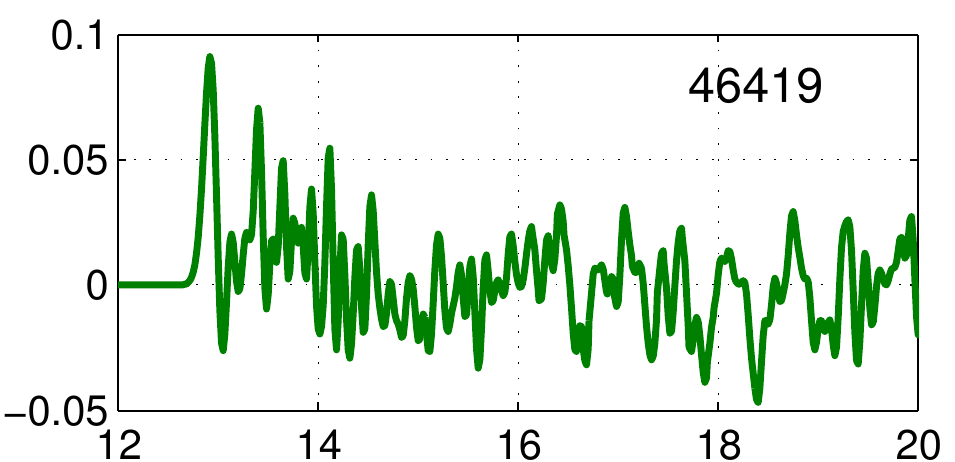} \\
	\includegraphics[width=0.5\textwidth]{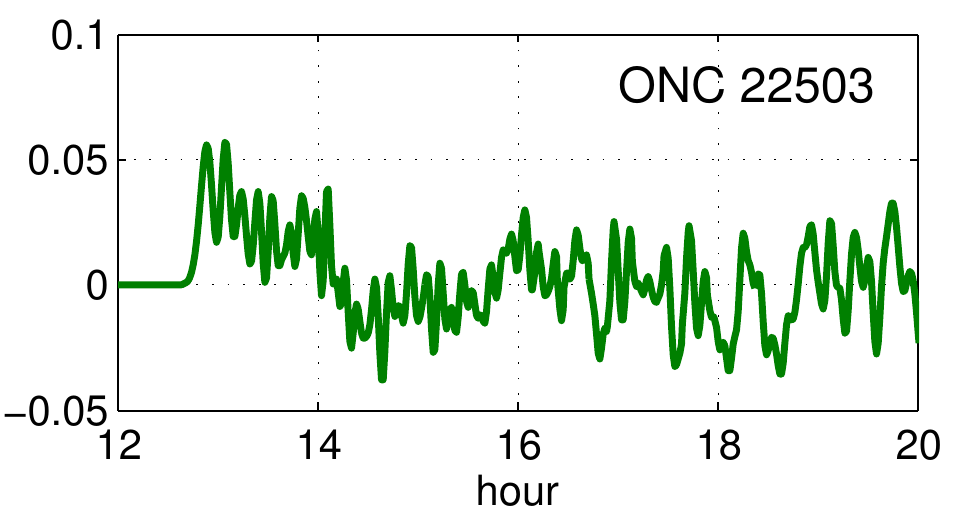} 
	\includegraphics[width=0.5\textwidth]{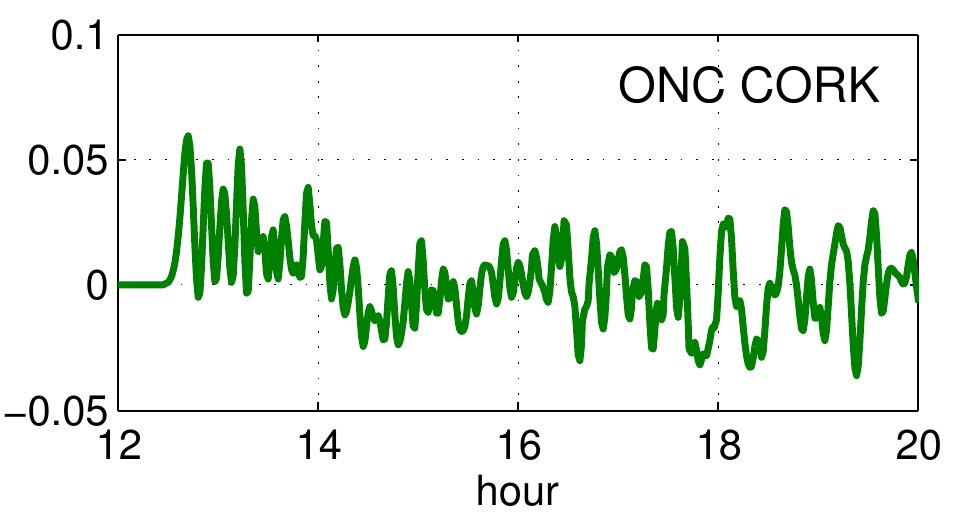} 
\end{tabular}
	\caption{Time histories of surface elevation (m) at DART 32412 and at offshore stations along North America (their PRFs to 32412, dimensionless), computed by propagating a narrow initial deformation centered on 32412, as described in \ref{math}. } 
	\label{responses2010}
\end{center}
\end{figure}

\begin{figure}[ht]
\begin{center}
\begin{tabular}{cc}
	\includegraphics[width=0.5\textwidth]{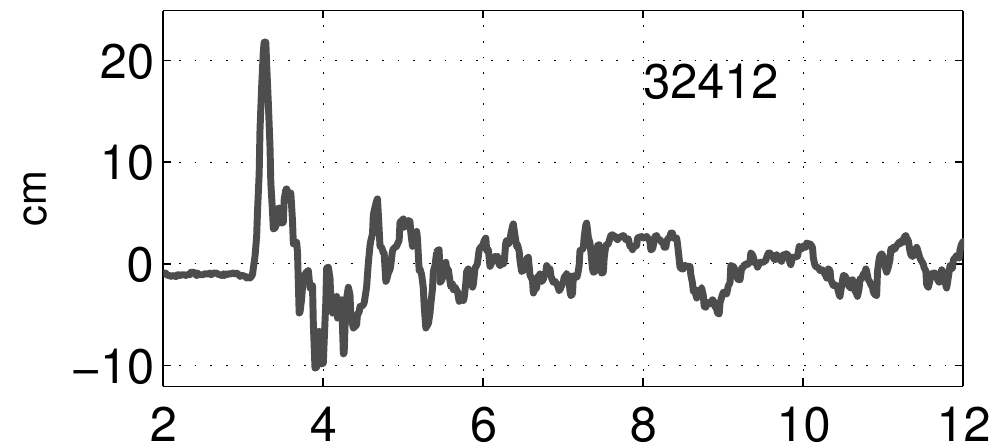} 
	\includegraphics[width=0.5\textwidth]{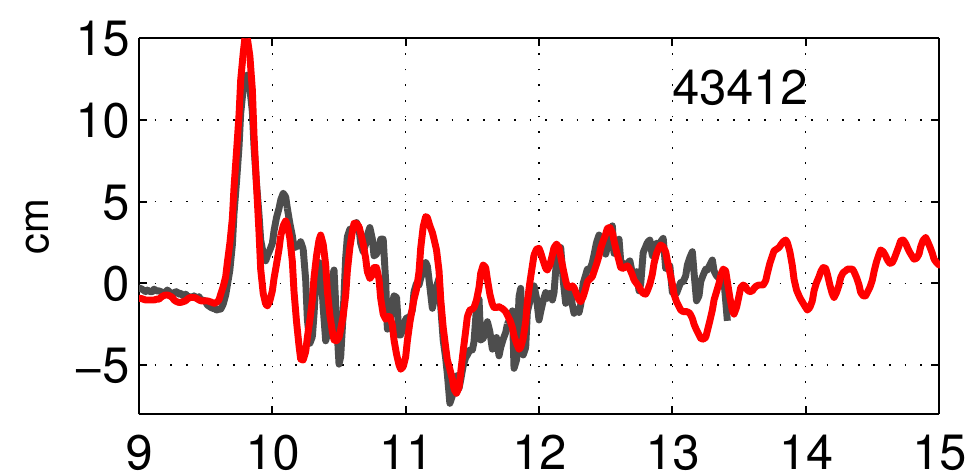} \\
	\includegraphics[width=0.5\textwidth]{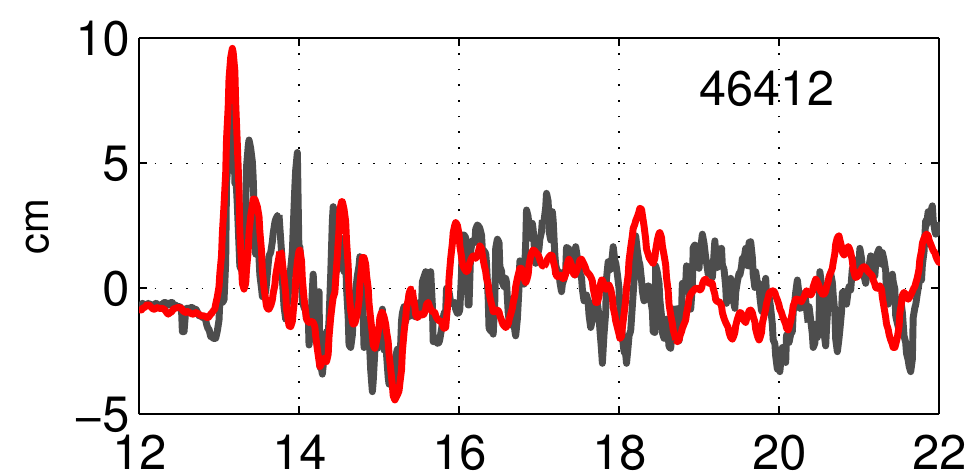} 
	\includegraphics[width=0.5\textwidth]{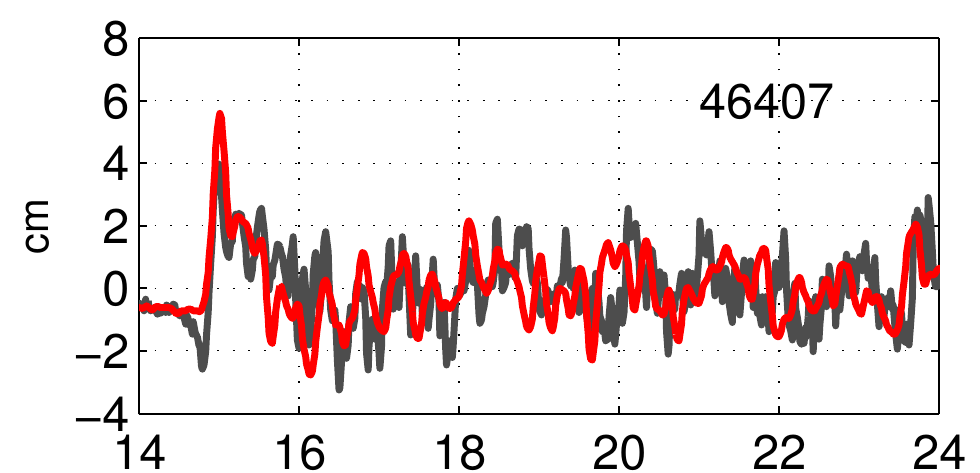} \\
	\includegraphics[width=0.5\textwidth]{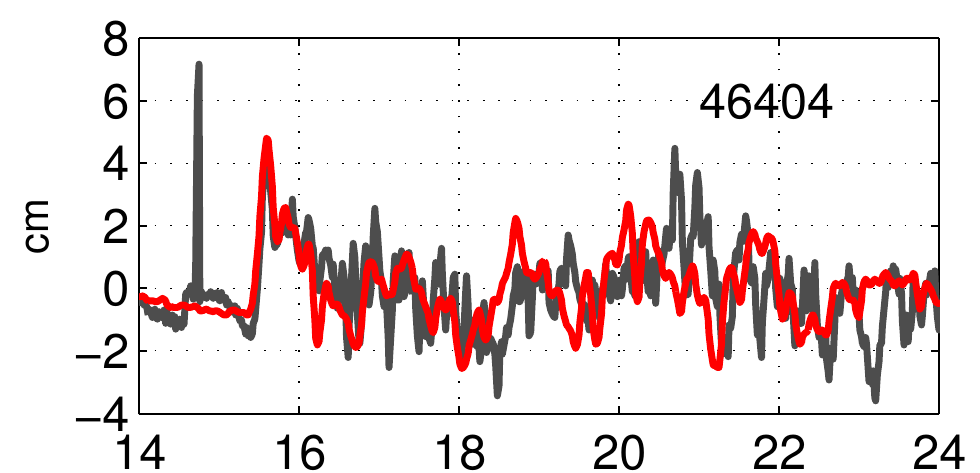} 
	\includegraphics[width=0.5\textwidth]{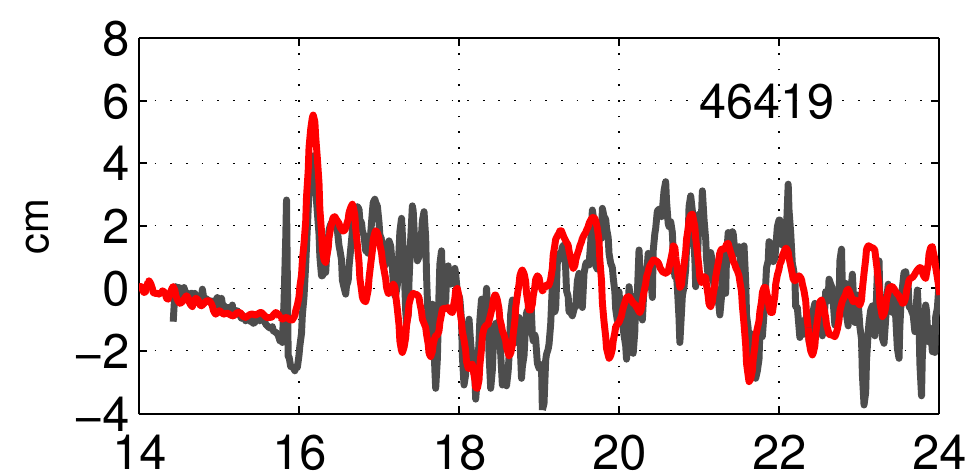} \\
	\includegraphics[width=0.5\textwidth]{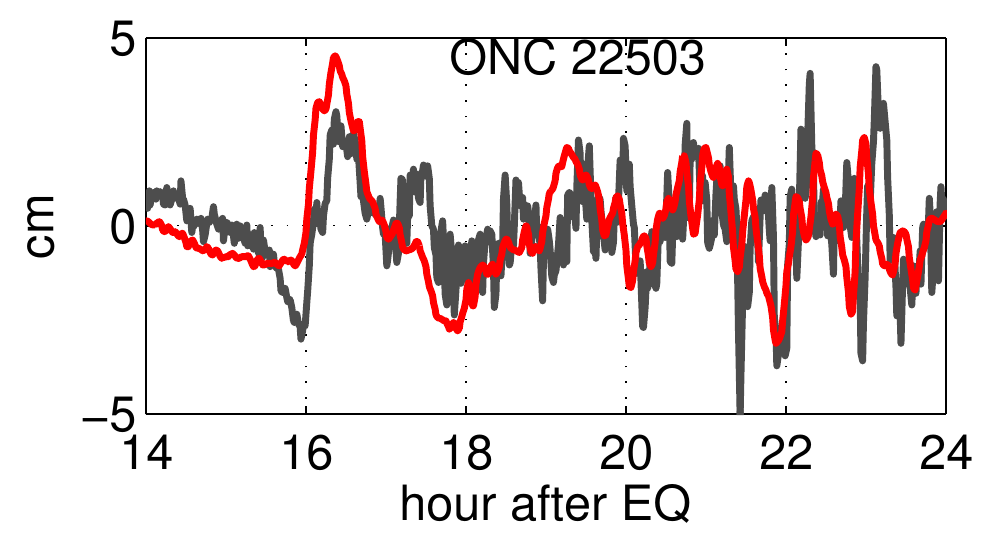} 
	\includegraphics[width=0.5\textwidth]{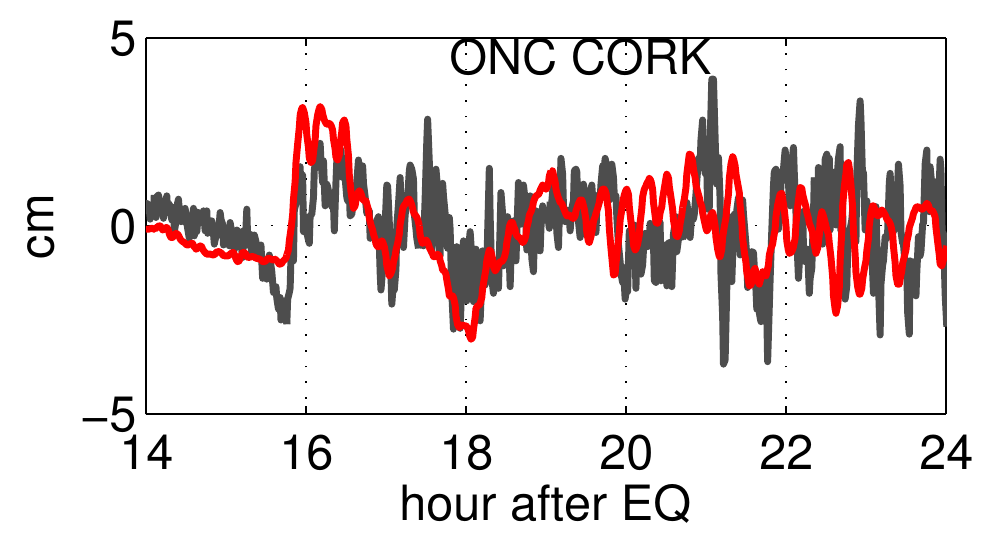} 
\end{tabular}
	\caption{Gray: observations of the 2010 Chile tsunami; Red/Black: Predictions at offshore stations along North America obtained as a response to the sea level variations at station 32412 off Peru. See Fig. \ref{mapCh} for the stations' locations. Time delay up to 6 min was applied to the hindcasts.}
	\label{gages2010}
\end{center}
\end{figure}

All simulations in this section were carried out with Cliffs using ETOPO bathymetry at 2 arc-min resolution. To compute the wavefront shape $\Gamma$, a small initial arc drawn around the EQ epicenter\footnote{a compact (point-like) source at the EQ epicenter can be used as well} was propagated numerically over realistic bathymetry until its hump reached the DART; the wavefront $\Gamma$ was defined by the coordinates under the maximal sea level elevation along the arc, as shown in Figure \ref{mapgamma}. As long as this initial arc is reasonably small, its radius and profile can be fairly arbitrary, since the wavefront shape at the detector is not sensitive to the details of the source function (the main premise of the method). In a land-free uniformly-deep sea, the latter premise would hold for a compact source with a small angular size as seen from the detector. 
In a real ocean, this premise is often facilitated by strong near-shore bathymetry, by narrowing the range of angles at which waves from a larger EQ area might arrive at the detector.

Figure \ref{mapCh} shows the sea surface in the 2010 Chile tsunami simulation at 4 h and 40 min after the tsunami origination. The tsunami simulation was initiated with an estimated source function for the 2010/02/27 event, produced with the SIFT system at the PMEL (\url{http://nctr.pmel.noaa.gov/chile20100227/}). Curve $\Gamma$, shown in Figure \ref{mapCh} by a red arc (after minor smoothing), conforms to the wavefront simulated with a realistic source function. As seen in the simulation, the radial structure of the wave changes at larger angles to the coastline, from a longer wavelength and a lower amplitude in the north-bound beam passing through 32412 to about twice shorter waves and a higher amplitude in the direction normal to the coastline. Consequently, the measurements at DART 32412 represent the actual wave time history only in the north-going tsunami wave beam, and permit forecasting only the sites along the American coasts north of the DART station.

The response computations were initialized with an initial deformation built upon $\Gamma$, as described in \ref{math} above. 
The resulting time-histories computed at the sites being forecasted are shown in Figure \ref{responses2010}. The initial deformation leaves a pulse-like signal at 32412, 2 min long until the first zero crossing (Figure \ref{responses2010}, top left), half of which corresponds to the forward-traveling wave. The solutions at other stations represent the stations' responses to DART 32412. 

The predictions shown in Figure \ref{gages2010} are obtained as convolutions of each site's response (shown in a corresponding pane in Figure \ref{responses2010}) and the tsunami measurements at station 32412 (shown in Figure \ref{gages2010}, top left). Consistent agreement between the predictions and the observations suggests that for any tsunami originating in Chile, its future time history anywhere by the coast of North America can be directly derived from the tsunami's record at the station 32412 (offshore Peru). 

Chilean-born tsunamis typically excite edge waves on the adjacent continental shelf, which sustain wave radiation for many hours \citep{geist2013, catalan2015}.
A relevant feature of the response formalism is its ability to account for any part of the wave train passing the detector at any time during the event, including the later waves resulting from local resonances, edge waves, and other shelf phenomena.
Owing to this feature, the resulting predictions are accurate for 10 hours or longer. 

\section{Case study: Predicting the 2011 Tohoku tsunami around the Pacific with three DARTs near Japan}
\label{tohoku}

\begin{figure}[ht]
\centering
	\resizebox{1.1\textwidth}{!} %
		{\includegraphics{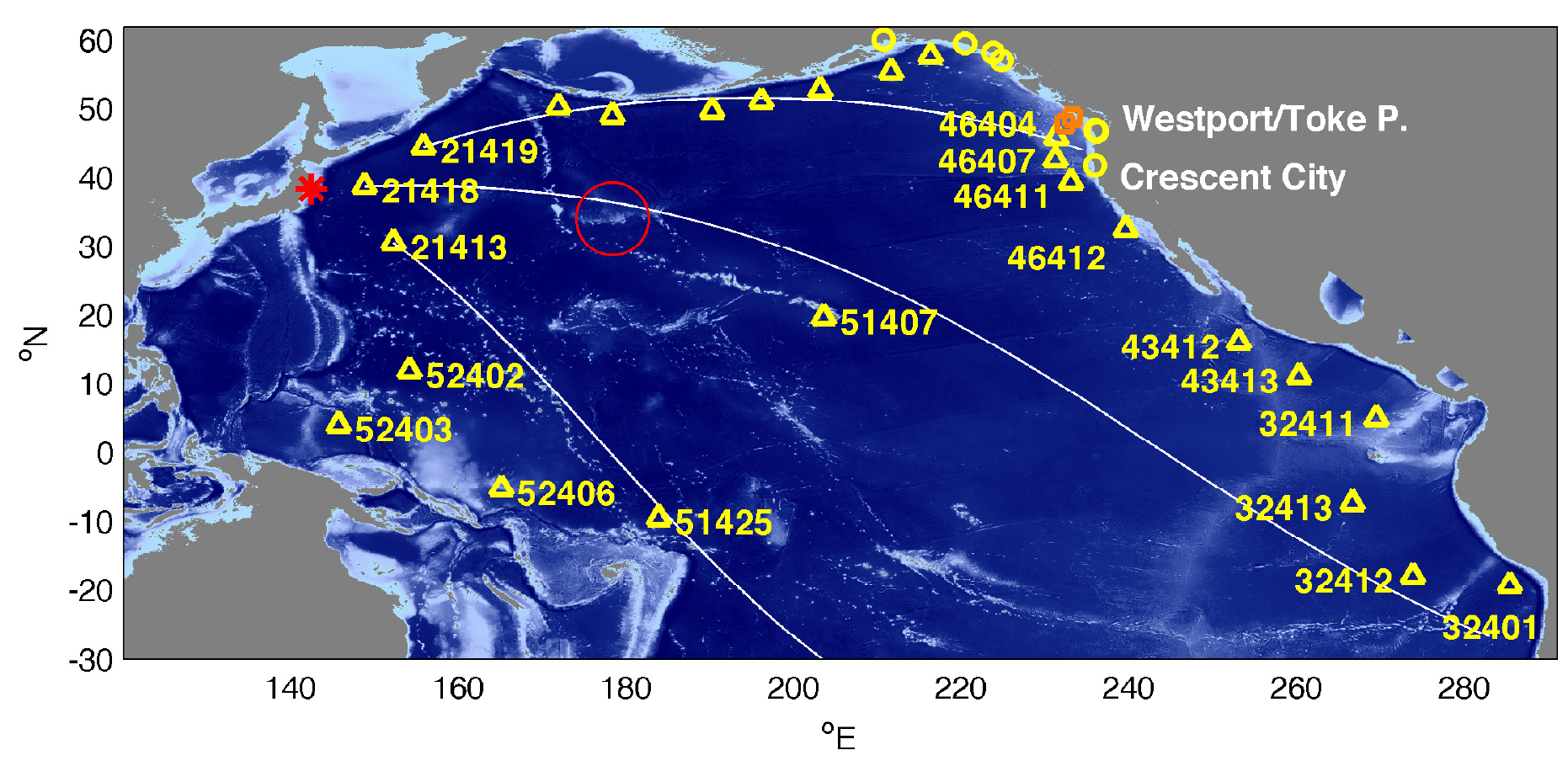}} 
	\caption{Triangles: DARTs; unassigned Aleutian and Alaskan DARTs (west to east) are 21415, 21414, 46408, 46402, 46403, 46409, and 46410; squares: two ONC BPRs offshore Vancouver Island; circles - coastal tide gages (left to right) at Seward, Yakutat, Elfin Cove, and Sitka in Alaska, Westport and Toke Point (unresolved in the map) in Washington, and Crescent City in California. 
Star (red/black) - the 2011/03/11 EQ epicenter. While/black lines show approximate tsunami paths from three DARTs near Japan, drawn  as great arcs normal to the wavefronts at the DARTs. Mellish Seamount is circled with red/gray. }
	\label{Tohmap}
\end{figure} 
Different from the Chilean example, tsunami waves generated by the Tohoku-2011 EQ arrived at the US West Coast following multiple routes. The direct wave train travelled along the Aleutians, whereas some later waves arrived to the West Coast refracted by sea mounts and underwater ridges in the mid-ocean \cite[]{tang2012}. Predicting this tsunami even at a single site might therefore require to use DARTs covering different tsunami routes. Here, combining responses to multiple detectors as described in section \ref{Nmath} is applied to hindcasting the 2011 Tohoku tsunami Pacific-wide, by predicting it from either two (21413 and 21419) or three (21413, 21418, and 21419) stations near the source area (see Figure \ref{Tohmap}). 

\subsection{Initial shapes}

\begin{figure}[ht]
\begin{center}
	\resizebox{\textwidth}{!} %
		{\includegraphics{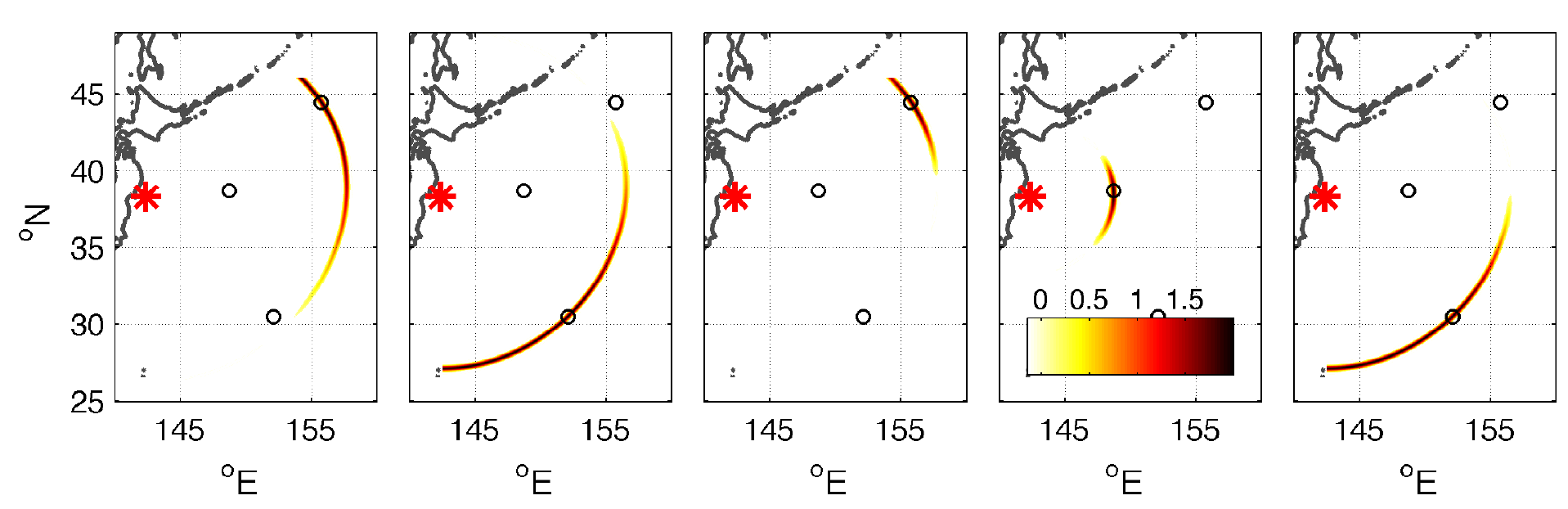}} 
	\caption{Initial surface deformations used to compute PRFs to the respesctive DARTs in the 2011 Tohoku event, in the 2-dart configuration (two left shapes) and in the 3-dart configuration (3 right shapes). Circles - DARTs 21413, 21418, 21419 (south to north). Star - EQ epicenter. Colorscale - initial surface displacement (m).}
	\label{Tohpulses}
	\end{center}
\end{figure} 
Responses at a site to initial shapes displayed in Figure \ref{Tohpulses} in the left two panes define the PRFs of the site to a 2-DART set of stations 21413 and 21419. Responses at a site to initial shapes displayed in Figure \ref{Tohpulses} in the right three panes define the PRFs of the site to a 3-DART set of 21413, 21418, and 21419.  
Each initial shape is given by equation \eqref{shape}; its centerline represents a segment of a leading front of a wave emanating from a point source in the epicenter of the 2011 Tohoku EQ, computed as in section \ref{chile}.  Linear azimuthal interpolation was applied to the initial shape at 21418 on both sides, at 21413 on the north side, and at 21419 on the south side. With all three detectors located in deep water seaside from the Japan trench, simple projections of the detectors onto the wavefronts were used to delimit the wavefront segments, with no ray tracing techniques employed in this particular case.
The initial shape passing through 21413 is uniform in azimuthal direction in its south-west section, and terminated by the Bonin Island; and the initial shape through 21419 is uniform at its north end and terminated by the Kuril Trench (Figure \ref{Tohpulses}). Each initial problem was simulated Pacific-wide for 27.5 hours, to obtain PRFs of various sites. The shortest, 7.5-hour-long PRF was computed for DART 32401 offshore central Chile, to where a pulse from 21418 (the station closest to Japan) travels 20 hours. The wavefront shapes and the PRFs were computed with Cliffs at 2 arc-min resolution with the ETOPO bathymetry.

\subsection{Pulse Response Functions}

\begin{figure}[ht]
\begin{center}
\begin{tabular}{c}
	\includegraphics[width=\textwidth]{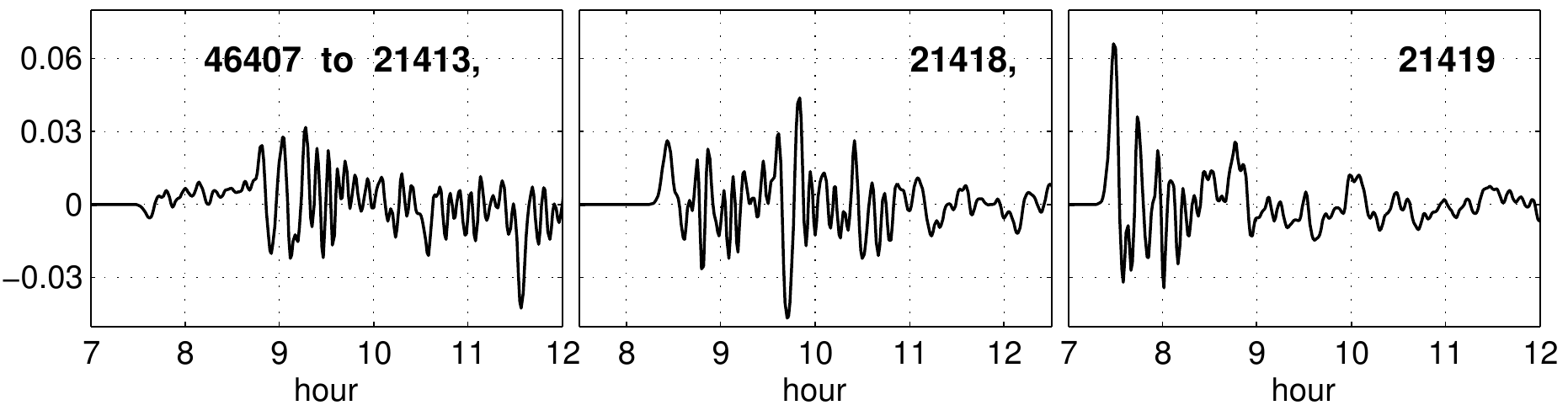} \\
	\includegraphics[width=\textwidth]{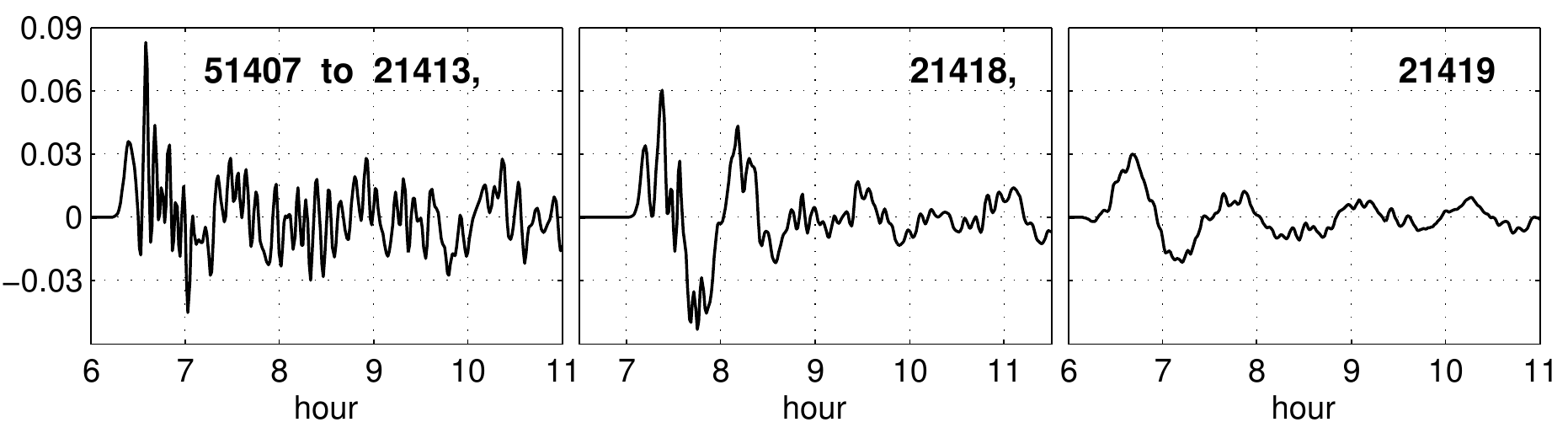} 
\end{tabular}
	\caption{PRFs of DARTs 46407 (top three plots) and 51407 (bottom three plots) to 21413, 21418, and 21419 in a wave assembled from wave beams passing through these three DARTs. Each PRF timing is counted from the moment when an initial pulse started its propagation from the corresponding detector.}
	\label{resps3d}
\end{center}
\end{figure}
Each PRF records transformations which the tsunami will experience between the detector and the site due to interactions with the complex bathymetry. As an example, we briefly examine PRFs at station 46407 off of the mid-Oregon coast and at station 51407 in Hawaii to each DART in the 3-DART set, shown in Figure \ref{resps3d}. The 46407's PRFs tell us that the first waves are mostly determined by the signal at 21419, and therefore arrived by the northern (along the Aleutians) route. Pulse response to 21413 becomes noticeable only in the later part, which indicates waves re-directed to the site by scattering. About 1.5 h after the pulse arrival, the PRFs to both 21413 and especially 21418 exhibit an intense signal at a distinct frequency, which indicates a strong bathymetric feature (Mellish Seamount, as follows). Likewise, the observed Tohoku tsunami signal at 46407 also contains intense waves between 1.5 and 2 h after the tsunami arrival, which Tang et al (2012) identified as a ray refracted from the Mellish Seamount. The 51407's PRFs are remarkable in how little of high frequency variance (typically dominating a PRF of a deep-sea site) is present in the pulse responses to 21418 and especially to 21419. As seen in the latter PRF, only a nearly harmonic wave with a 72-min period  arrives to Hawaii from the 21419's wave beam traveling along the Aleutians. It is possible that the underwater ridges connecting Hawaii and Aleutians serve as a wave guide at this particular wavelength (this hypothesis has not yet been evaluated). 
\subsection{Predictions}
\begin{figure}[ht]
\begin{center}
\begin{tabular}{c}
	\includegraphics[width=\textwidth]{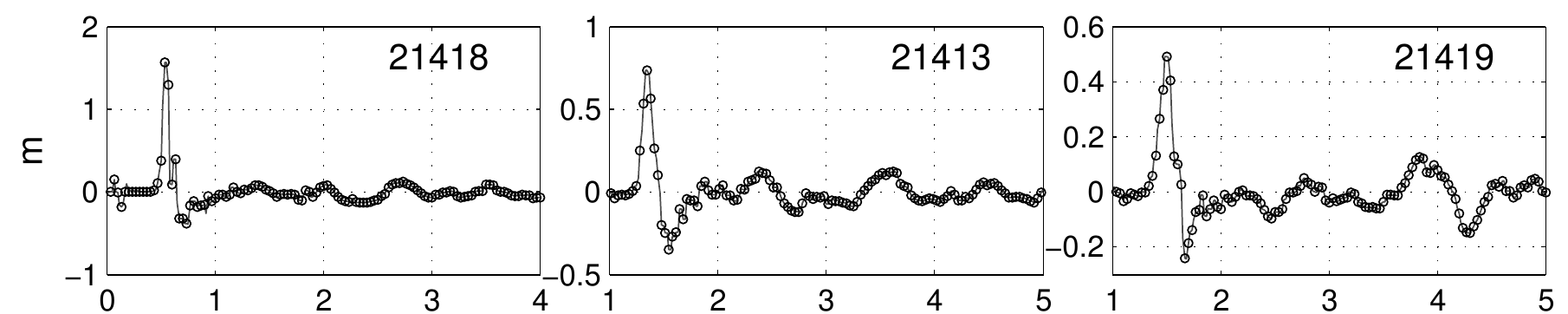} \\
	\includegraphics[width=\textwidth]{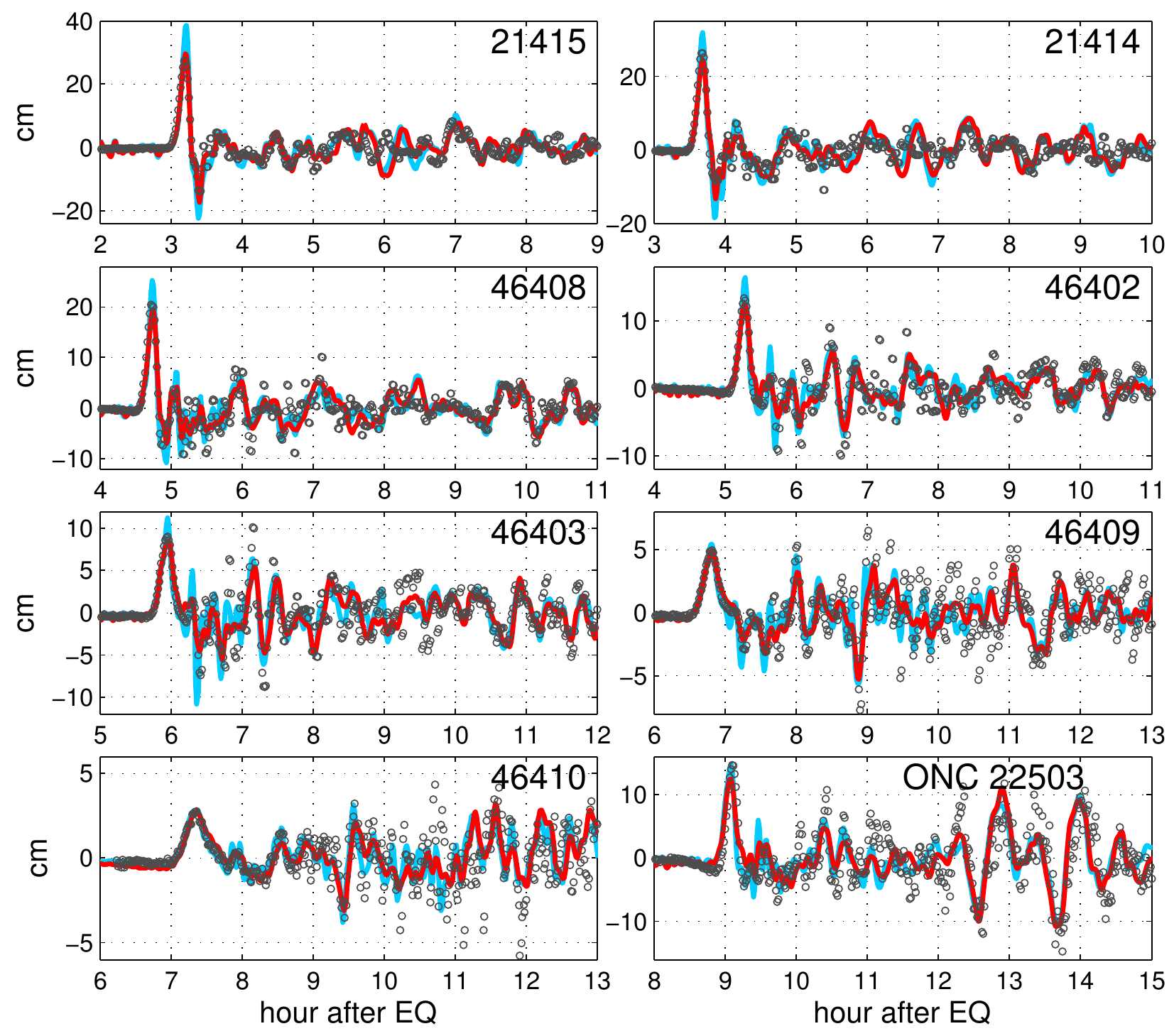} 
\end{tabular}
	\caption{Top row: the 2011 Tohoku tsunami measurements at stations 21418, 21413, 21419 near Japan, circles denote readings with a 2-min interval used for the hindcasting. Lower rows: tsunami observations (black circles) and predicted response to two detectors 21413/21419 (red) and to three detectors 21413/21418/21419 (blue) at deep sea stations along Aleutials, Alaska, and northern North America.}
	\label{TohDD1}
\end{center}
\end{figure}

\begin{figure}[ht]
\begin{center}
	\resizebox{\textwidth}{!} %
		{\includegraphics{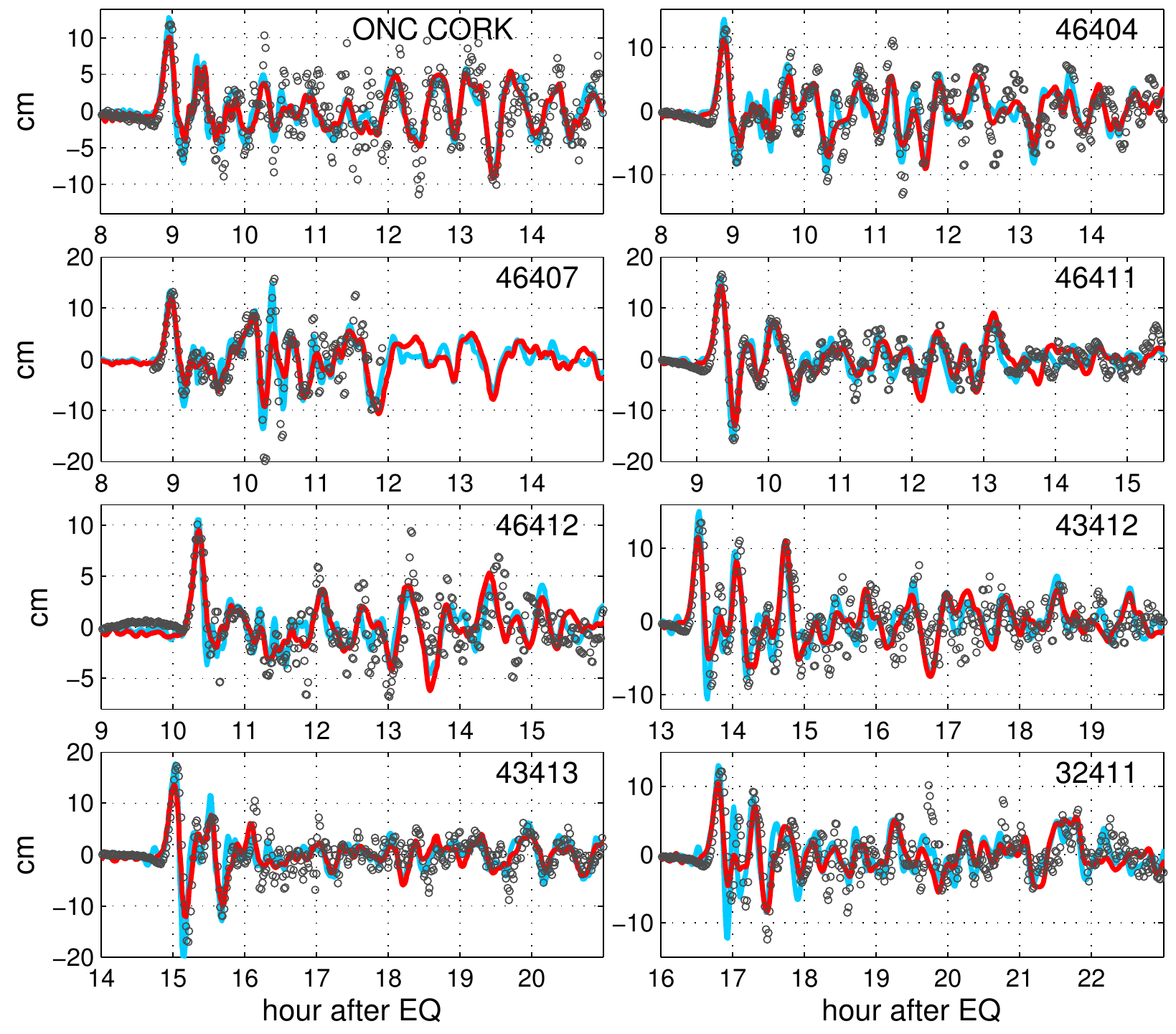}}
	\caption{The 2011 Tohoku tsunami observations (black circles) and predicted response to a 2-dart set (red) and a 3-dart set (blue) at more DART stations by the US West Coast and Central America.}
	\label{TohDD2}
	\end{center}
\end{figure}

\begin{figure}[ht]
\begin{center}
	\resizebox{\textwidth}{!} %
		{\includegraphics{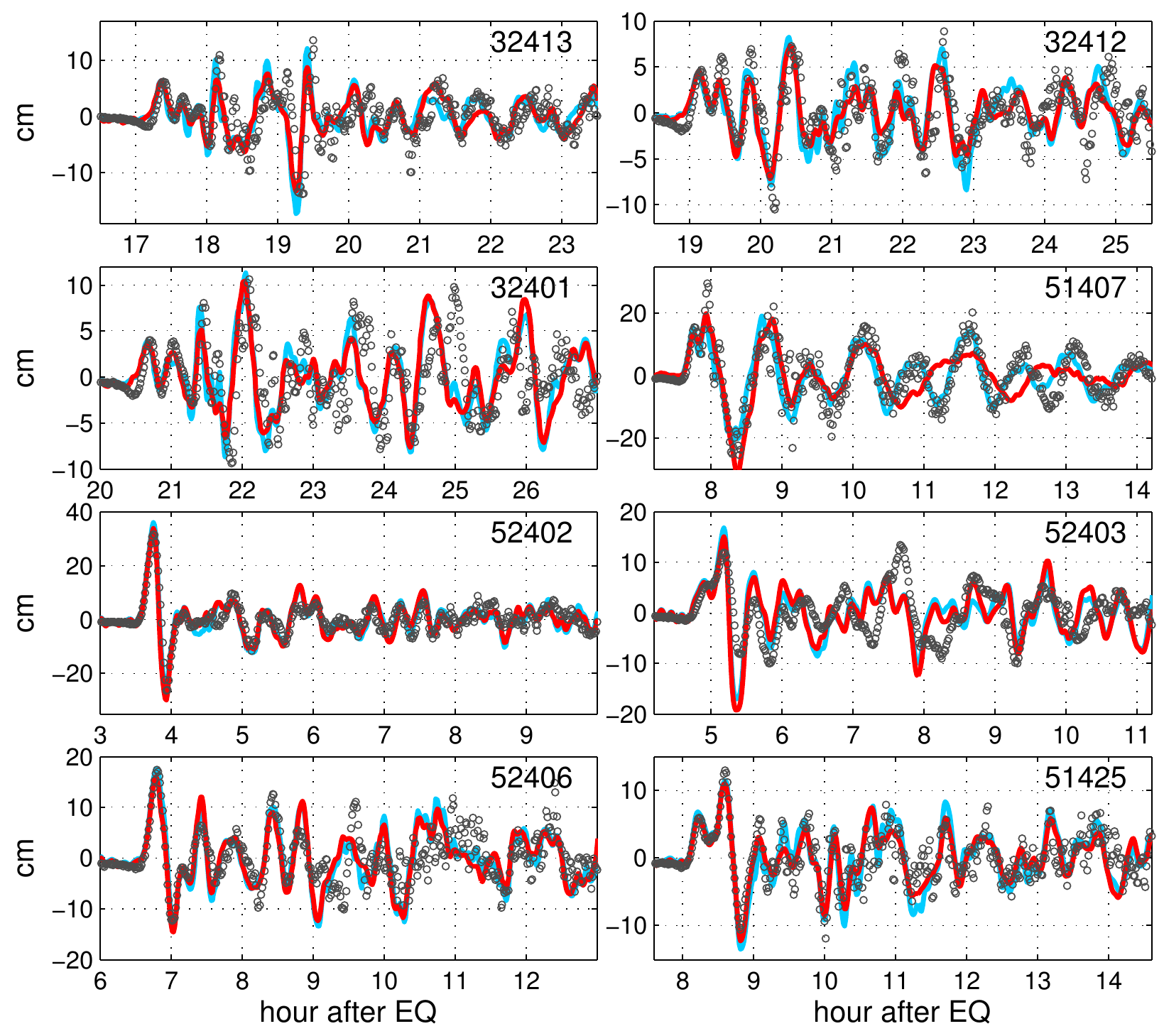}} 
	\caption{The 2011 Tohoku tsunami observations (black circles) and predicted response to a 2-dart set (red) and a 3-dart set (blue) at DART stations offshore South America, Hawaii, and  south-west Pacific.}
	\label{TohDD3}
	\end{center}
\end{figure}

Given the pre-computed PRFs, the prediction at each site is obtained as a sum \eqref{ss} of the responses to the wave motion registered at each DART in a set, each response being a convolution of a DART signal and a corresponding PRF.
Figure \ref{TohDD1}, top row, displays the 2011 Tohoku tsunami measurements (de-tided) at stations 21413, 21418, and 21419 (the detectors), sampled at a 2-min interval, used to produce the predictions at 24 deep-water stations around the Pacific. These stations also recorded the Tohoku tsunami (Figure \ref{Tohmap}), where it arrived between 1.6 and 20 hours after it had passed the detectors. The resulting predictions computed as a combined response to the 2-detector set and to the 3-detector set, vs. the observed sea level variations, are shown in Figures \ref{TohDD1}-\ref{TohDD3}. By visual comparison with the SIFT-made predictions \cite[]{tang2012} with an estimated source inferred from the measurements at two DARTs 21418 and 21401 (in the direction of 21419, out of service since 07/2014) and at three DARTs 21418, 21401, and 21413, the hindcasts computed with the response formalism display a similar accuracy. 

More DARTs around the source detect more variability along the wavefront. 
In particular, as seen from the records at the three detectors (Figure \ref{TohDD1}, top row), the tsunami has a larger wave height and a shorter wavelength in the central beam passing through 21418, while it has a smaller wave height and a longer wavelength in the side beams through 21413 and 21419. Thereby the predictions made with the 2-DART set do not resolve this central wave beam.
Nevertheless, both the 2-DART and the 3-DART predictions have captured the major signal variance at each location.
The 2-DART predictions fit the observations better at five Aleutians DARTs, which receive their wave, shaped by the Kurils and Aleutian trenches, directly from 21419 (see Figure \ref{Tohmap}). 
Contribution from 21413 in the Aleutians is negligible, due to a very different beam direction.
In the 3-DART configuration, however, Aleutian sites receive a signal from 21418 introduced by azimuthal interpolation. This signal is responsible for the false short peaks on top of correctly predicted waves. 

At all other locations, the 3-DART predictions provide a better approximation to observations due to resolving the more intense shorter-wave central beam missed by the 2-DART predictions. 
In particular, only the 3-DART prediction reproduced waves arriving at 46407 between hr 10 and 11 after the EQ. These are waves refracted from the Mellish Seamount \cite[]{tang2012}. The 3-DART prediction continues to fit the observations at 51407, whereas the 
2-DART prediction starts to deviate from the recorded signal 3 hr after the arrival. Both predictions at 51407 miss a narrow peak near the top of the first wave. As noted by Tang et al (2012), station 51407 is located only 60 km offshore the Island of Hawaii, so adequate modeling of wave dynamics at this station might require higher spatial resolution. The only site where the predictions noticeably deviated from the record is station 52403, located farthest west  -- too far on the ``edge" of the wave assembled from the three wave beams (see Figure \ref{Tohmap}). \\

It has been found that numerical forecasts systematically underestimate tsunami arrival time in the far field, which was explained by the neglect of seawater and seafloor compressibility and some other factors \citep{tsai, watada2014}. The propagation time errors embedded in the computed PRFs might potentially degrade the combined responses,
should these errors be different among the site's responses to different detectors.
In the Tohoku case, the difference among the propagation time errors from a detector to the site in different tracks is not expected to exceed 1-3 min, which, given the much longer tsunami period, is not likely to significantly affect the results. Nevertheless, since the response formalism with multiple detectors is potentially sensitive to phase errors, the PRF computations in this section were performed using a depth correction method suggested by Wang (2015). Almost perfect timing between the predictions and the observations at all sites (no time shifts were applied to any time history in Figures \ref{TohDD1}-\ref{TohDD3}) suggests that the propagation time errors have been largely eliminated. Without the depth correction, the predicted arrival times are up to 15 min early; while the predicted wave heights with and without depth correction differ very mildly for this event.   
\section{Case study: Same Pulse Response Functions, different events}
\label{btest}

Pulse Response Functions connect a forecasted site and the deep-sea detectors, and depend on the location of the tsunami originating area. At the same time, PRFs do not depend on the tsunami source structure. No information about an EQ geometry except a location of its epicenter were used for deriving PRFs exercised in this work.  We therefore expect that the same PRFs used for forecasting, for instance, the 2011 Tohoku tsunami, can be used for forecasting another tsunami originating in the same area -- though how large of an area the same responses can serve varies from region to region and is a subject for a separate study. This proposition is illustrated here by applying the above PRFs used with the 2011 Tohoku tsunami (centroid at $142.373^oE$ and $38.297^oN$, at 29.0 km depth) for predicting two other, artificial tsunamis. The tsunamis were generated, accordingly, (1) by a hypothetical $M_w$ 8.0 trust fault EQ caused by a uniform 3.6 m slip of a 169 km x 47 km plate ($(strike, dip, rake)=(187,15,90)$) with a centroid at  $143.25^oE$ and $39.0^oN$ at 20 km depth, 109 km north of the epicenter of the 2011 Tohoku EQ; and (2) by a hypothetical $M_w$ 8.3 normal fault EQ caused by a uniform 5.6 m slip of a 247 km x 58 km plate ($(strike, dip, rake)=(206,45,-90)$) with a centroid at $144.5^oE$ and $37.5^oN$ at 26 km depth, 207 km south of the epicenter of the 2011 Tohoku EQ. For both hypothetical events, a shear modulus of 44.1 GPa is assumed.

\begin{figure}[ht]
\begin{center}
\begin{tabular}{c}
	\includegraphics[width=\textwidth]{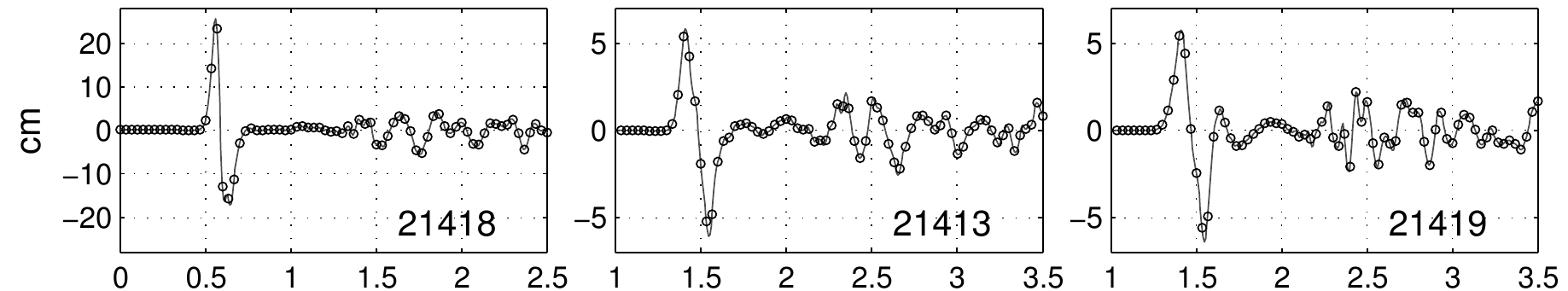} \\
	\includegraphics[width=\textwidth]{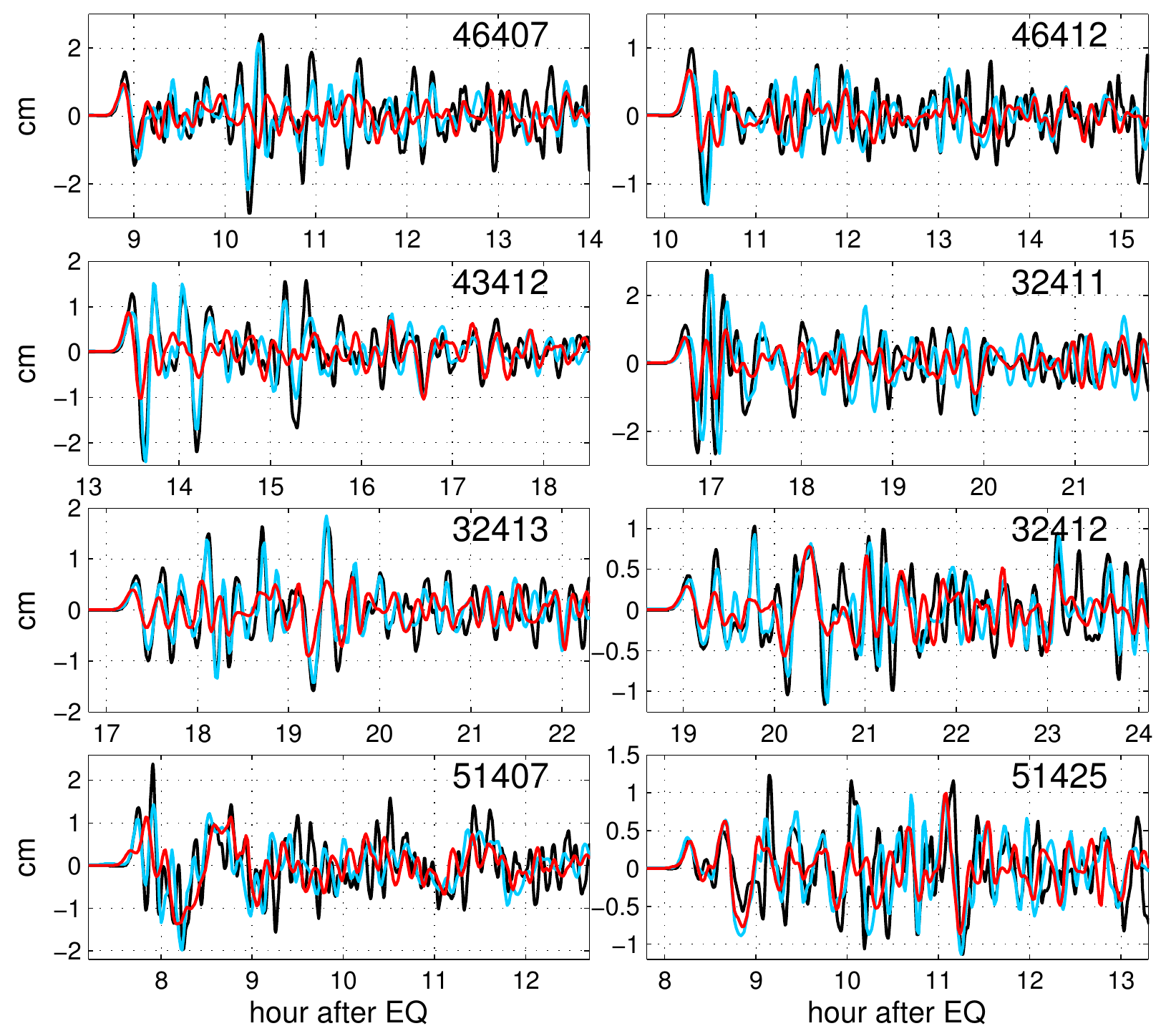} 
\end{tabular}
	\caption{Top row: the first artificial tsunami measurements at stations 21418, 21413, 21419; circles denote readings with a 2-min interval. Lower four rows: simulated observations (black) and predicted response to two detectors 21413/21419 (red) and to three detectors 21413/21418/21419 (blue) at selected DARTs by the US West Coast, Central America, South America, Hawaii (51407), and Samoa (51425).}
	\label{btest1}
\end{center}
\end{figure}

\begin{figure}[ht]
\begin{center}
\begin{tabular}{c}
	\includegraphics[width=\textwidth]{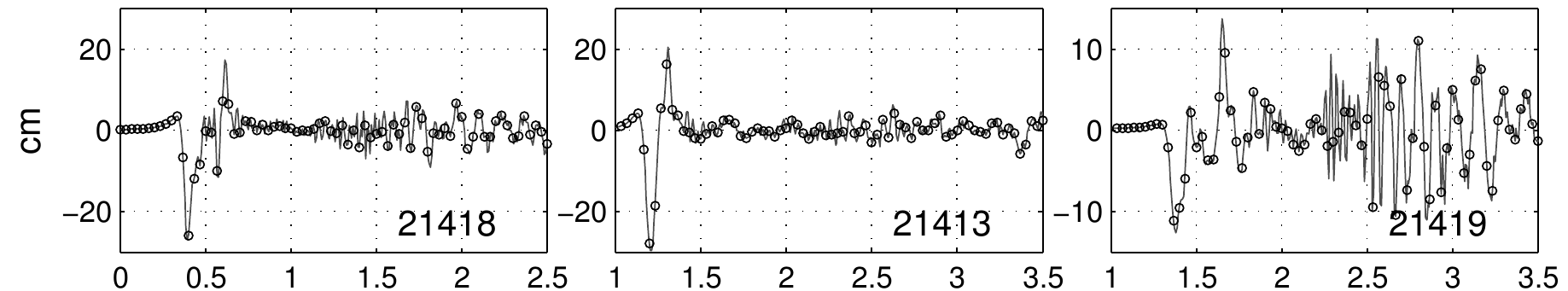} \\
	\includegraphics[width=\textwidth]{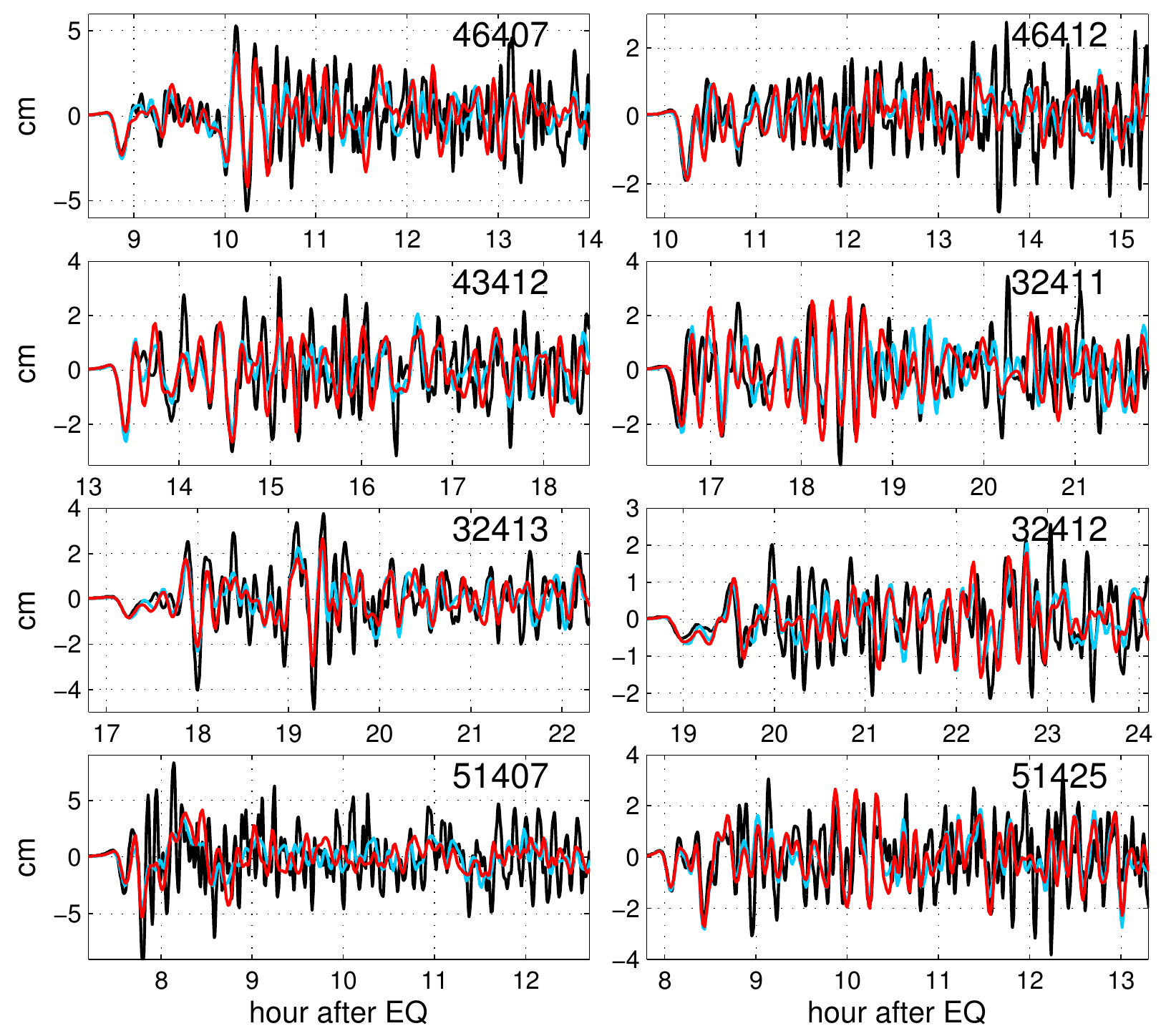} 
\end{tabular}
	\caption{Top row: the second artificial tsunami measurements at stations 21418, 21413, 21419; circles denote readings with a 2-min interval. Lower four rows: simulated observations (black) and predicted response to the station pair (red) and trio (blue) at selected DARTs by east Pacific coasts, Hawaii (51407), and Samoa (51425).}
	\label{btest2}
\end{center}
\end{figure}

The artificial tsunamis were propagated from the source to the DART stations using the RIFT model  \cite[]{wang2012} developed and operated at the PTWC. The simulations were carried out at a 4 arc-min resolution. The bathymetric grid was derived from the GEBCO 30-arc-sec data, and the ocean-depth correction method of Wang (2015) was applied to reduce the tsunami travel time errors. Compared to the 2011 Tohoku tsunami, the artificial tsunamis are composed of much shorter waves. For instance, the duration of the full leading wave of the Tohoku tsunami at 21413/21419 is about 30 min vs. 20 min (Figure \ref{btest1}) for the first artificial tsunami and 12 min for the second artificial tsunami, followed by even shorter waves (Figure \ref{btest2}). 

The first artificial tsunami is similar to the Tohoku tsunami in that it has a higher-amplitude shorter-wave central beam and lower-amplitude longer-wave side beams.
Results of predicting this tsunami at a number of stations selected to represent different regions (the US West Coast, Central America, South America, Hawaii, and Samoa) are shown in Figure \ref{btest1}. The difference between the predictions made with the 2-detector set and the 3-detector set is very appealing: the 2-detector predictions miss a lot of signal variance, while the 3-detector predictions fit the simulated ``observations" almost perfectly. A better azimuthal coverage provided by more detectors was essential, since it had permitted to reproduce variability along the wavefront. 
Thereby, three main factors set the lower limit on the duration of a wave\footnote{Herein, term ``wave" refers to an individual wave, positive or negative, singled out by two consecutive zeros in a record of the sea surface displacement, and not to a harmonic component in the record.}  successfully forecasted with the response formalism:
\begin{itemize}
\item
azimuthal resolution, roughly determined by a number of detectors around the source area;
\item
temporal sampling interval at the detectors ($\Delta t_0=2$ min, throughout this work), prescribing the width of the initial shape for computing the PRFs and the shortest possible period ($2\Delta t_0$) in the detected signal;
\item
numerical properties of the model(s) in the short (low-resolved) wave range.
\end{itemize}
Results of predicting the second, even shorter-period tsunami shown in Figure \ref{btest2} are largely affected by these three factors.
This tsunami has marginally (for a 2-min sampling rate) high frequency, and its records at the three detectors do not indicate the presence of an energetic central beam. That is, this beam was either absent, or was not detected. Predictions made with 2 DARTs and 3 DARTs are very similar, closely reproduce the longer-period waves, and miss the higher frequency variance. 
In particular, the later short waves poorly resolved at the detectors with an adopted 2-min sampling rate are also missing from the predictions. 
In addition, RIFT and Cliffs models differ in numerical dissipation in the short-wave range. RIFT model utilizes a loss-less difference scheme, whereas Cliffs uses the same numerical stencil as the MOST model which dissipates low-resolved waves \cite[]{diffdisp}. Therefore, PRFs computed with Cliffs might contain less high-frequency energy than the solutions of the RIFT model (the truth is probably somewhere in between). Waves with a longer duration (10+ min), however, have been accurately predicted at all locations. 

\clearpage
\section{Forecasting at tide gages} 
\label{sgages}

\begin{figure}[ht]
\begin{center}
\begin{tabular}{c}
	\includegraphics[width=0.8\textwidth]{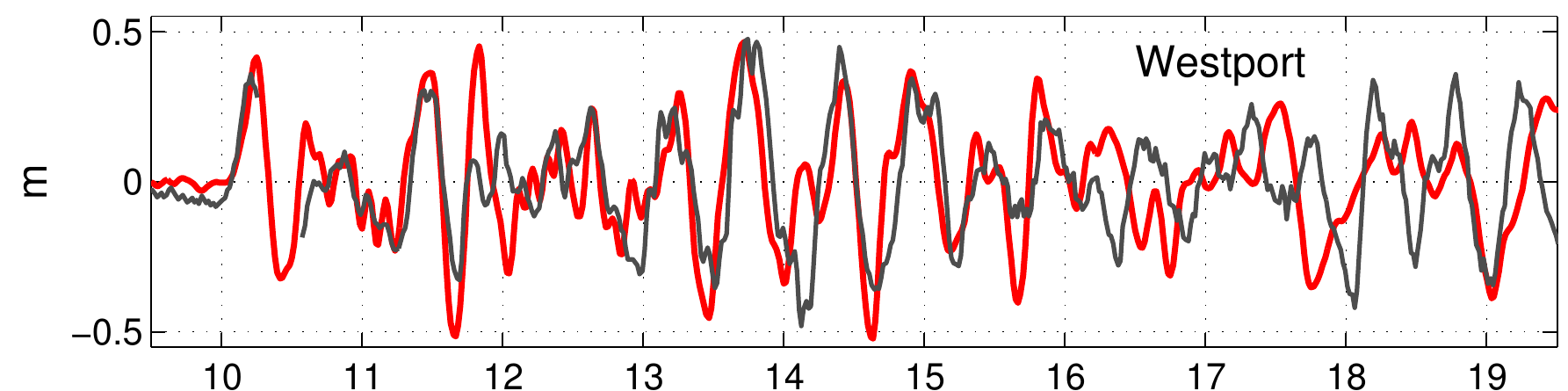} \\
	\includegraphics[width=0.8\textwidth]{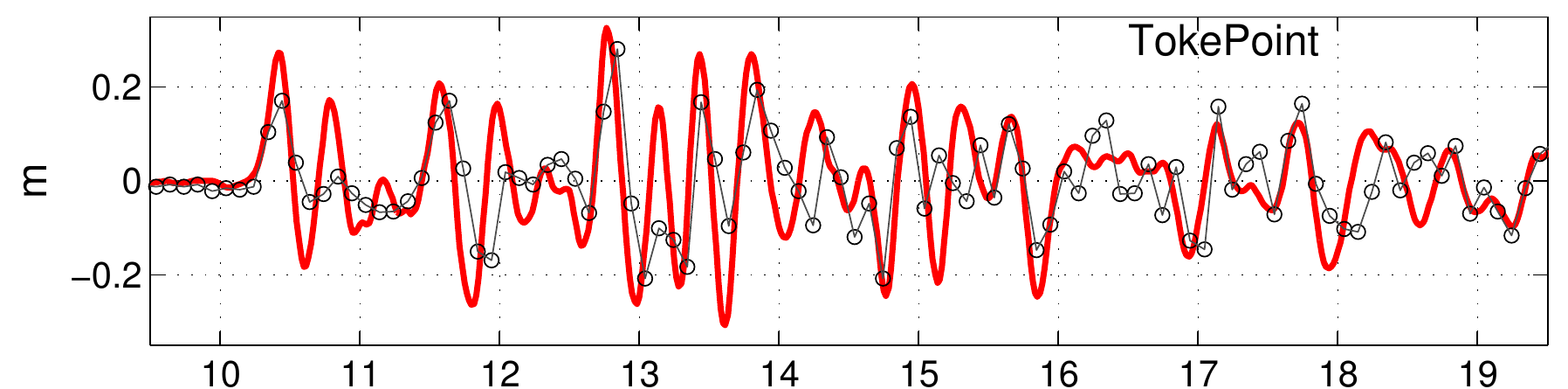} \\
	\includegraphics[width=0.8\textwidth]{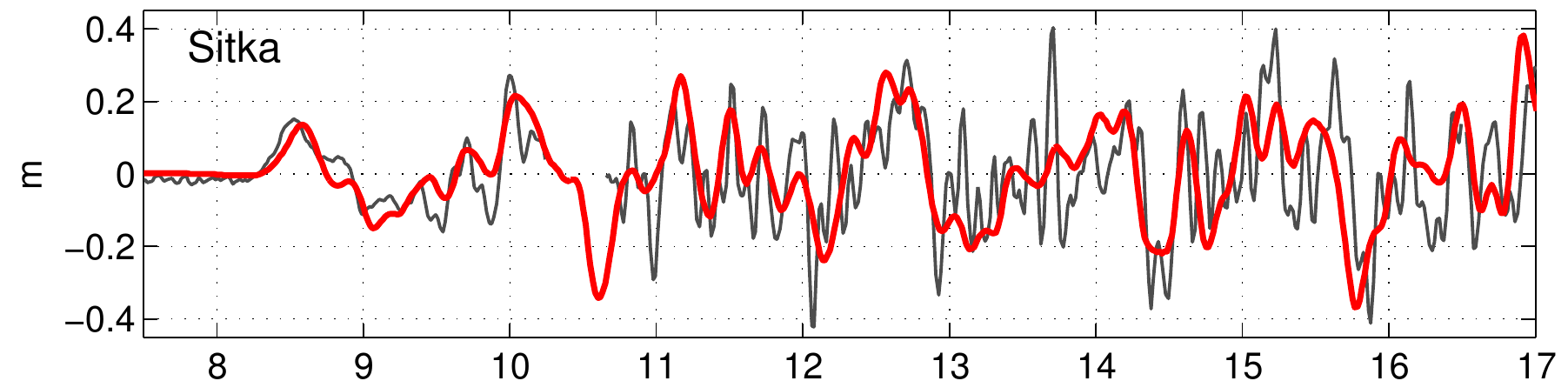} \\  
	\includegraphics[width=0.8\textwidth]{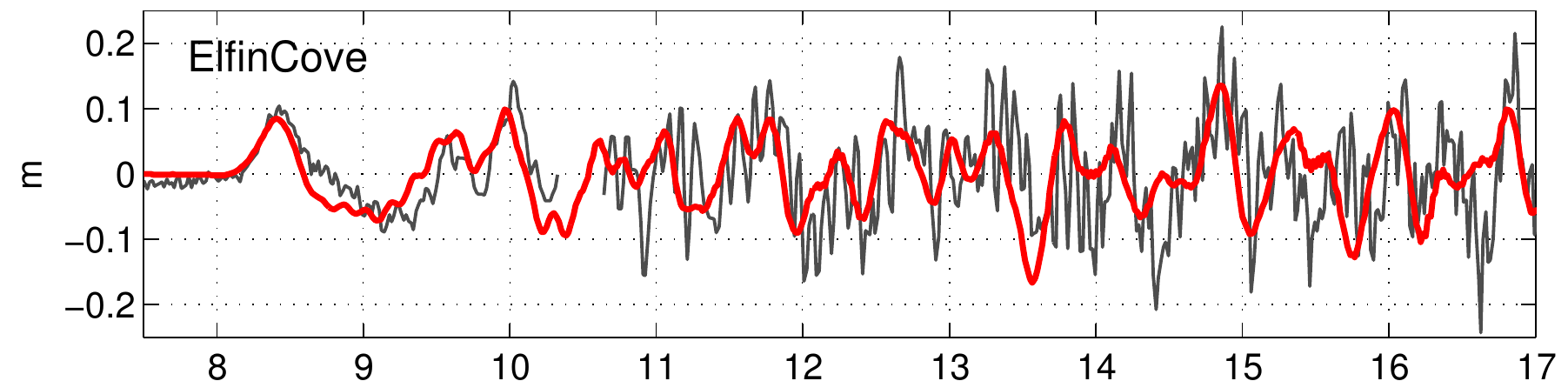} \\
	\includegraphics[width=0.8\textwidth]{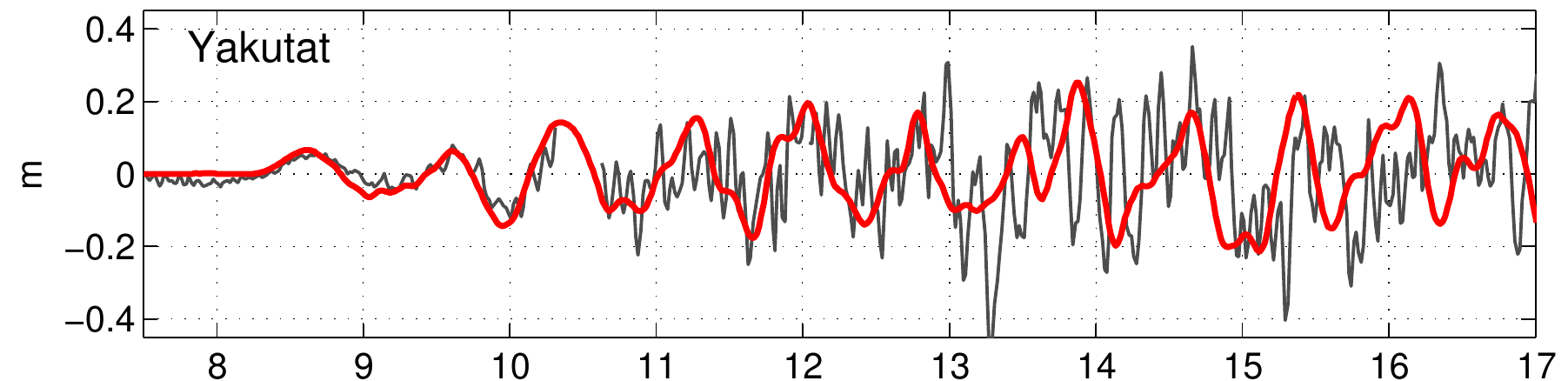} \\
	\includegraphics[width=0.8\textwidth]{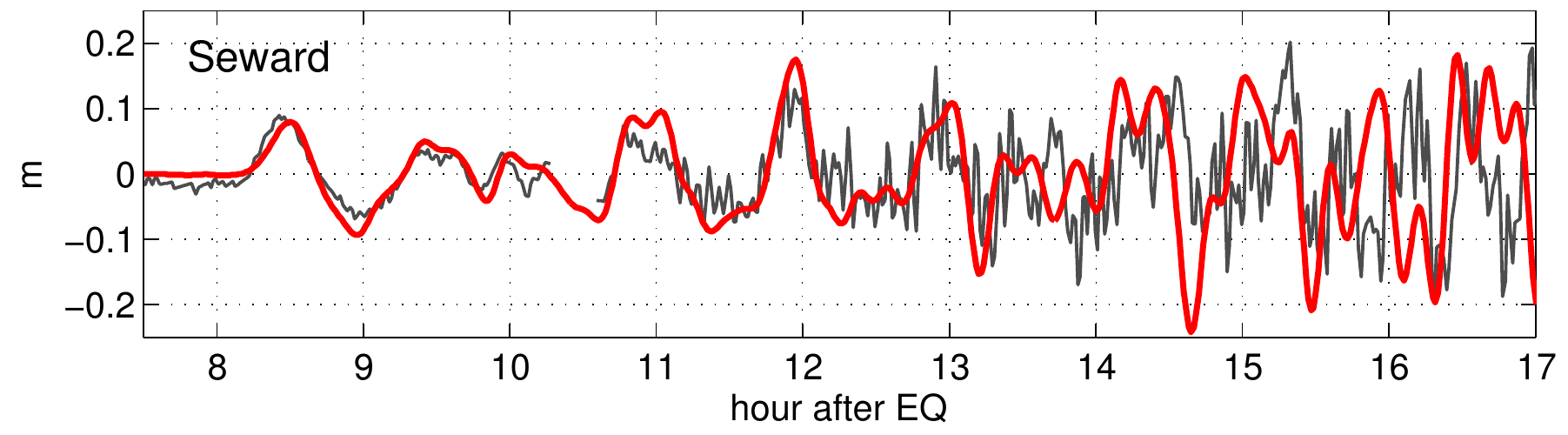} 
\end{tabular}
	\caption{Red: Predictions of the 2011-Tohoku tsunami as a combined response at a coastal site to the 3-dart set (21413, 21418, 21419); Black:  observations sampled at a 1-min interval (solid line) and at a 6-min interval (circles).}
	\label{gages2011}
\end{center}
\end{figure}

To compute a PRF of a coastal gage to a DART, an initial pulse was propagated from the DART to the gage, using a nested-grid system with the increasing spatial resolution focusing on the gage. Figure \ref{gages2011} presents 3-dart (21413, 21418, 21419) hindcasts vs observations of the 2011 Tohoku tsunami at tide gages in Westport and Toke Point in Washington, and Sitka, Elfin Cove, Yakutat, and Seward in Alaska. 
Local responses to the DARTs were computed with Alaska Forecast Model \cite[]{nicolsky} running within ATOM modeling package for the locations in Alaska, and with Cliffs for the locations in Washington, using 3-5 levels of nesting starting with a 2 arc-min spaced grid in the deep ocean and refining the resolution to 1-3 arc-sec in the inner-most grids containing the gages. The grids are based on high-resolution digital elevation models developed by the NOAA's National Centers for Environmental Information (NCEI) to support tsunami modeling and coastal inundation mapping. Manning friction coefficient $n=0.03$  $s \cdot m^{-1/3}$ was used.
In the presented cases, the response method showed almost perfect reconstruction of the longer period variance. For shorter waves, the reconstruction is less perfect. Predicting very short waves is probably beyond the capabilities of the shallow-water class of tsunami models.
The application of the response formalism is limited by the condition of linearity of the propagation tract from the detector to the site (hereafter, a wave is termed `linear' if its amplitude at a point of interest is linearly related to the wave amplitude in the deep sea, and termed `non-linear' otherwise).
Thus the response method might not directly apply to forecasting large waves in shallow bays and harbors or to predicting the details of land inundation,
but can predict relatively small waves, as in the above examples.
Moreover, the response method can be modified to allow forecasting some non-linear environments, in particular, moderately big waves in some bays and harbors, as described in Sections \ref{sgages1}-\ref{sgages2} below. 

\subsection{Equations for a non-linear resonator}
\label{sgages1}
An equation describing a forced oscillation of a typical linear resonator reads:
\begin{equation}
s_{tt}+ \sigma s_t+ \omega^2 s=forcing
\label{r1}
\end{equation}
with constant parameters $\sigma$ and $\omega$ that represent a dissipation rate and a resonance frequency; a subscript denotes derivation. 
This equation has an explicit solution for the state variable $s$ (the forecast): 
\begin{equation}
s=G(\sigma, \omega) \otimes forcing,
\label{r2}
\end{equation}
where the impulse response $G$, analogous to PRF, depends on the system's parameters. 
A non-linear resonator is often described by a similar equation\footnote{typically, \eqref{r1} is merely an approximation of \eqref{r3} when the  oscillations are small}, where the system's parameters depend on the system's state
\begin{equation}
s_{tt}+ \sigma(s, s_t) s_t+ \omega^2(s, s_t) s=forcing.
\label{r3}
\end{equation}
For moderately large oscillations still following an approximately harmonic law, we substitute \eqref{r3} with 
\begin{equation}
s_{tt}+ \sigma(||s||, ||s_t||) s_t+ \omega^2(||s||, ||s_t||) s=forcing,
\label{r4}
\end{equation}
where $|| \cdot ||$ denotes some norm of a variable, also referred to as a signal level. In doing so, we are treating the non-linear resonator as a linear one with the parameters specific for the signal level in an unknown solution. Using \eqref{r2} yields:
\begin{equation}
s=G \left(\sigma(||s||, ||s_t||), \omega(||s||, ||s_t||) \right) \otimes forcing.
\label{r5}
\end{equation}
This expression presents another form of equation \eqref{r4}. We want to solve \eqref{r5} without having to compute $G$ in the solution process, but using a pre-computed $G$ set instead. We note that \eqref{r5} is equivalent to a system of two equations with respect to two functions $s$ and $\xi$:
\begin{eqnarray}
\label{r6}
s&=&G\left(\sigma(||\xi||, ||\xi_t||), \omega(||\xi||, ||\xi_t||) \right) \otimes forcing \\
\label{r7}
||s||&=&||\xi||
\end{eqnarray}
Let $\xi_k$, $k=1, \dots, K$ be a pre-defined set of $K$ functions such that $||\xi_1||<||\xi_2||< \dots <||\xi_K||$, and the set $||\xi_k||$ samples the range of targeted signal levels in $s$. Let $s_k$ be a forecast computed with \eqref{r6} for $\xi=\xi_k$. If there exists a unique index $j$ for which $||s_j|| \approx ||\xi_j||$, then $s=s_j$ is the sought solution. 
Next, this approach is elaborated for forecasting a harbor with a set of PRFs.

\subsection{Harbor forecasting with scaled PRFs}
\label{sgages2}

A harbor, as a resonator, is characterized by two parameters -- a resonance frequency and a Q-factor. Non-linearity affects, first of all, the dissipation rate: a more energetic wave meets a resonator with a lower Q-factor. 
As discussed in \ref{sgages1}, we attempt to treat a harbor as a linear resonator with the parameters specific for a signal level in the harbor. The term ``signal level" refers to a wave amplitude temporarily established in the harbor in response to wave forcing from the ocean. A harbor will be characterized by a set of PRFs, hereafter referred to as scaled PRFs, computed for a set of pulse heights $a$ at the deep-ocean detector(s) as
\begin{equation}
p_{a}(t)=\frac{1}{a}P_a(t),
\label{pa}
\end{equation}
where 
$P_a(t)$ is a numerical solution in the harbor to the fully-nonlinear SWE, initialized with a respective initial condition at the remote deep-ocean detector multiplied by factor $a$. Apparently, for $a_1<a_2< \dots <a_K$, 
\begin{equation}
||P_{a_1}||<||P_{a_2}||< \dots <||P_{a_K}||. 
\label{r10}
\end{equation}
A set $a$ for computing scaled PRFs is selected to provide a desired set of levels $||P_a||$ which sample the range of targeted signal levels in the harbor. 

\begin{SCfigure}
\centering
	\caption{Signal levels in a hypothetical harbor: in response to initial shapes with height $a$ (black dots), and in predictions of a hypothetical tsunami made with corresponding scaled PRFs (orange triangles); $x$-axis -- pulse height $a$. The ``right" prediction is circled with red.}
		\includegraphics[width=0.6\textwidth]{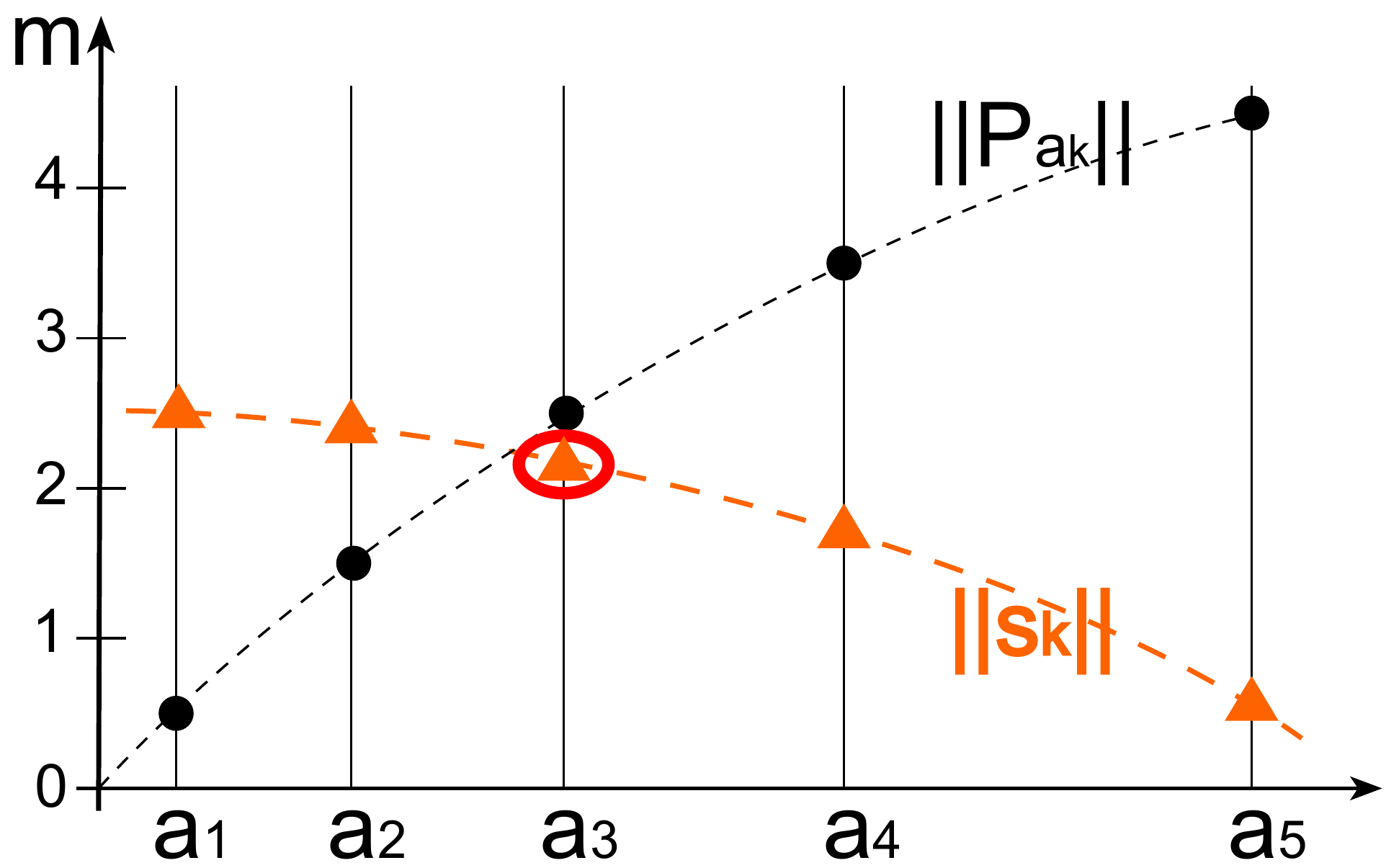}
	\label{scales}
\end{SCfigure}

For pulse heights $a$ exciting small waves in a harbor, the harbor's scaled PRFs $p_a(t)$ approach its linear limit $p(t)$ independent of $a$. The bulk of the energy loss for a small-amplitude wave is due to radiation out of the harbor rather than due to quadratic friction, so the harbor behaves as a linear system. For a larger wave amplitude, a scaled PRF $p_a(t)$ describes a resonator with larger frictional losses. A scaled PRF is therefore lower than the linear-limit PRF. The greater the $a$ value, the lower the scaled PRF. 
Therefore predictions $s_k$ produced with scaled PRFs satisfy 
\begin{equation}
||s_1||>||s_2||> \dots >||s_K||. 
\label{r11}
\end{equation}
Due to relations \eqref{r10} and \eqref{r11}, curves $||s_k||$ and $||P_{a_k}||$ are bound to intersect in a single point for any signal $s$ in the range $0 \le ||s|| \le ||P_{a_K}||$, as depicted in Figure \ref{scales}. Prediction $s_j$ closest to the intersection is therefore the sought forecast.

When forecasting with more than one DART, scaled PRF computations should use the common set of levels $||P_a||$, whereas       
an $a$ set can vary among the DARTs. Contributions from individual detectors in the sum \eqref{ss} should use scaled PRFs corresponding to the same $||P||$ level. Therefore, scaled PRFs can provide as many predictions as there are selected $||P||$ levels, regardless of the number of DARTs.

Prediction levels greater than  $||P_{a_K}||$ indicate very high waves outside the targeted range.
The response formalism won't be able to evaluate the height of such waves, but will clearly predict their occurrence. 

\subsection{Case study: Forecasting Crescent City, CA}
\label{sgages3}

\begin{figure}[ht]
\begin{center}
\begin{tabular}{c}
	\includegraphics[width=\textwidth]{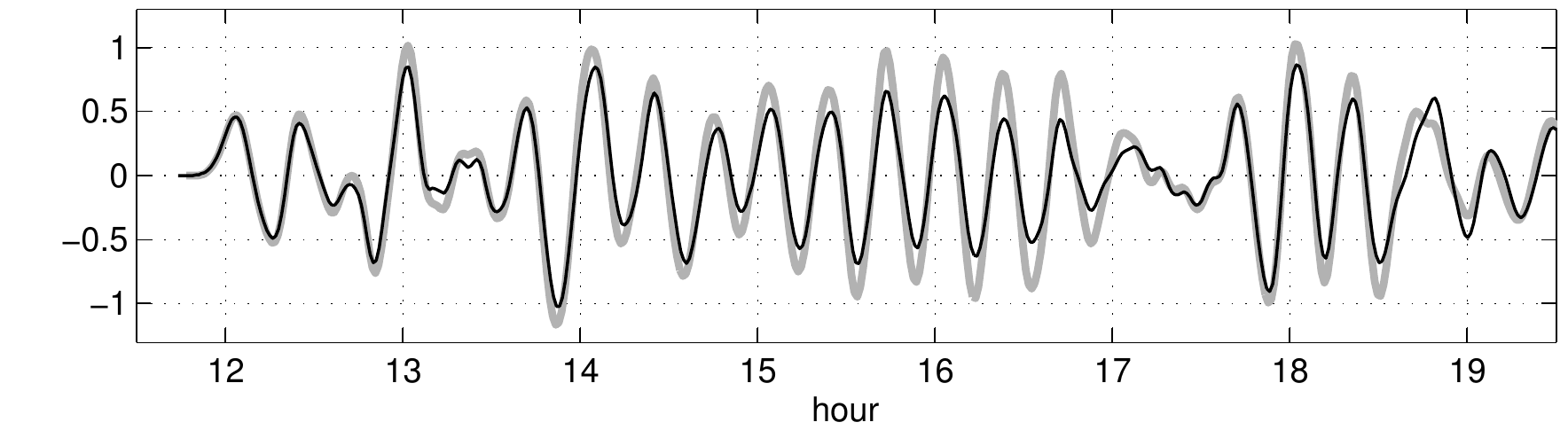} \\ 
	\includegraphics[width=\textwidth]{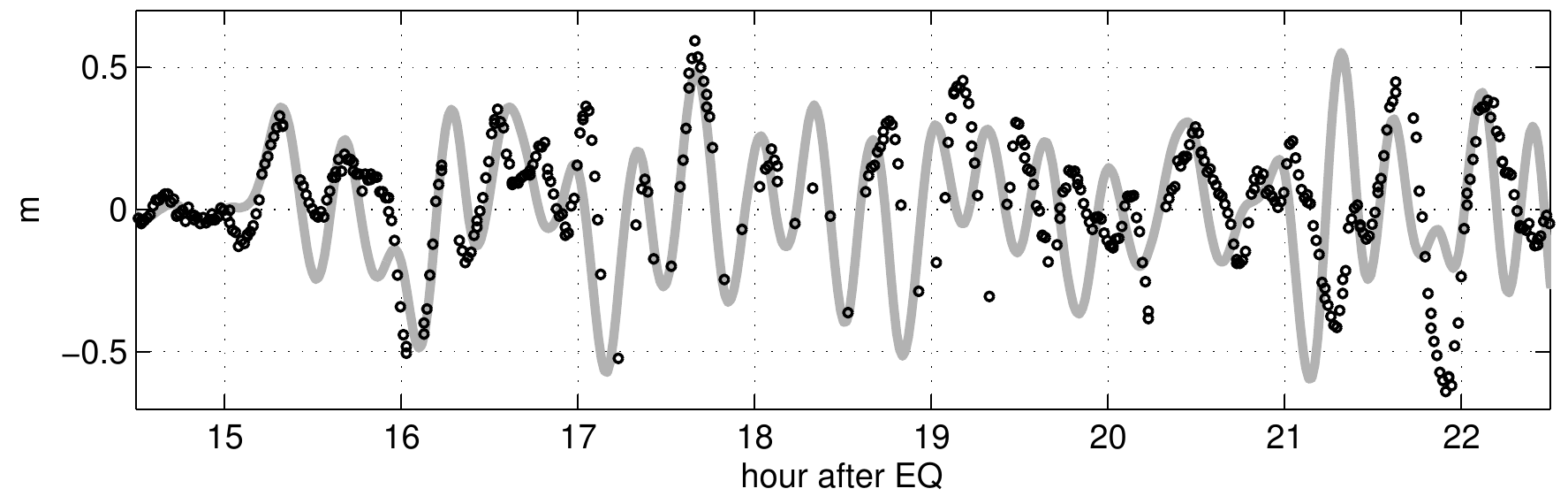}
\end{tabular}
	\caption{ Top: PRFs of the Crescent City tide gauge to DART 32412 computed with a $a=0.1$ m high pulse at the DART (gray) and $a=0.5$ m high pulse (black). Bottom: Prediction at Crescent City made with the 2010 Chilean tsunami recordings at station 32412 using a PRF for $a=0.5$ m
vs. the observations merged from 1-min sampled and 6-min sampled de-tided sea level measurements (black circles). A 6 min delay is applied to the PRFs.}
	\label{CC1}
	\end{center}
\end{figure}

In this manner, we'll attempt forecasting the Crescent City harbor in northern California, known for its extreme sensitivity to tsunamis \citep{wilson2013, xing2013}. Figure \ref{CC1}, top panel shows scaled PRFs of the NOAA/NOS tide gage in the harbor to DART 32412 for the origin area of the 2010/10/27 tsunami, computed for $a=0.1$ m (a proxy for the linear-limit PRF) and $a=0.5$ m. As anticipated, $|p_{0.1}| \ge |p_{0.5}|$, though appreciable difference is observed only 3+ hours after the tsunami arrival. Therefore, harbor's oscillations with waves under 0.5 m can be considered linear to weakly non-linear. Figure \ref{CC1}, the bottom panel shows the 2010/10/27 tsunami prediction computed with the tsunami record at 32412 and the scaled PRF $p_{0.5}$, selected so that a signal level in the resulting prediction matches a signal level in $0.5 \cdot p_{0.5}$ (m), while a prediction made with the linear-limit PRF starts to over-estimate after 3 h (not shown).

\begin{figure}[ht]
\begin{center}
	\resizebox{\textwidth}{!} %
		{\includegraphics{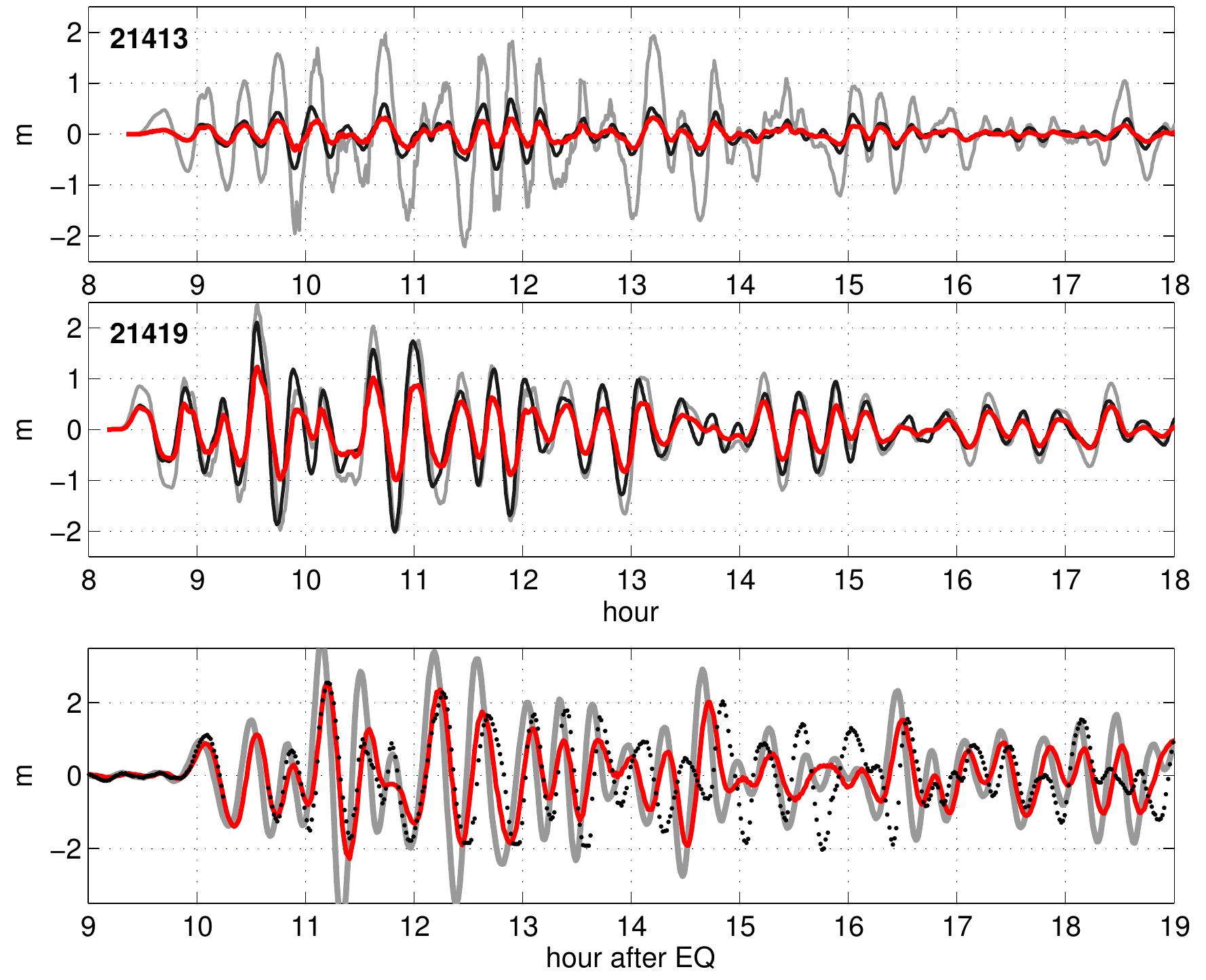}} 
	\caption{Top panel: responses at the Crescent City gage to a 6-m high pulse (gray) at station 21413, and the gage PRFs (black - the linear limit PRF, red - the scaled PRF, label 'm' does not apply). Middle panel: responses at the Crescent City gage to a 2-m high pulse (gray) at station 21419, and the gage PRFs (black - the linear limit PRF, red - the scaled PRF, label 'm' does not apply). Bottom panel: predictions of  the 2011 Tohoku tsunami at Crescent City as a combined response to two stations 21413 and 21419, using the linear-limit PRFs (gray) and the scaled PRFs (red), vs. the observations (black dots).}
	\label{CC}
	\end{center}
\end{figure} 

Unlike the 2010 event, the harbor's response in the 2011 Tohoku tsunami was strongly non-linear. Figure \ref{CC}  shows 2-DART hindcasts of the 2011 Tohoku event at the harbor's tide gage. The top two panels display responses $P_a(t)$ in m and dimensionless PRFs $p_a(t)$ of the gage to the DARTs 21413 and 21419. The linear-limit PRFs are drawn with black lines. Gray lines show responses $P_a(t)$ to a 6-m high initial shape at 21413 and a 2-m high shape at 21419, respectively. The amplitudes of $a=6$ m at 21413 and $a=2$ m at 21419 were selected to produce waves in the harbor up to 2 m in amplitude, in response to each initial shape. Red lines show the corresponding scaled PRFs, which are lower than the linear-limit PRFs over the later part of the signal, but coincide with them over the first period (non-linearity takes time to develop). The scaled PRFs yield a prediction (red in the bottom panel) varying up to 2 m in amplitude as the responses $P_a(t)$ to either DART do. Hence we have matched signal levels in a response to a scaled pulse at each DART and in a 2-DART prediction produced with the scaled PRFs. 
The prediction agrees well with the observations (black dots, bottom panel), in spite of the non-linearity of the system. As expected, a prediction made with the linear-limit PRFs (gray in the bottom panel) significantly over-estimated the harbor's actual response for the more energetic part of the wave train. \\

\section{Case study: Forecast timing}
\label{wtime}

\begin{figure}[ht]
\begin{center}
\begin{tabular}{c}
	\includegraphics[width=0.6\textwidth]{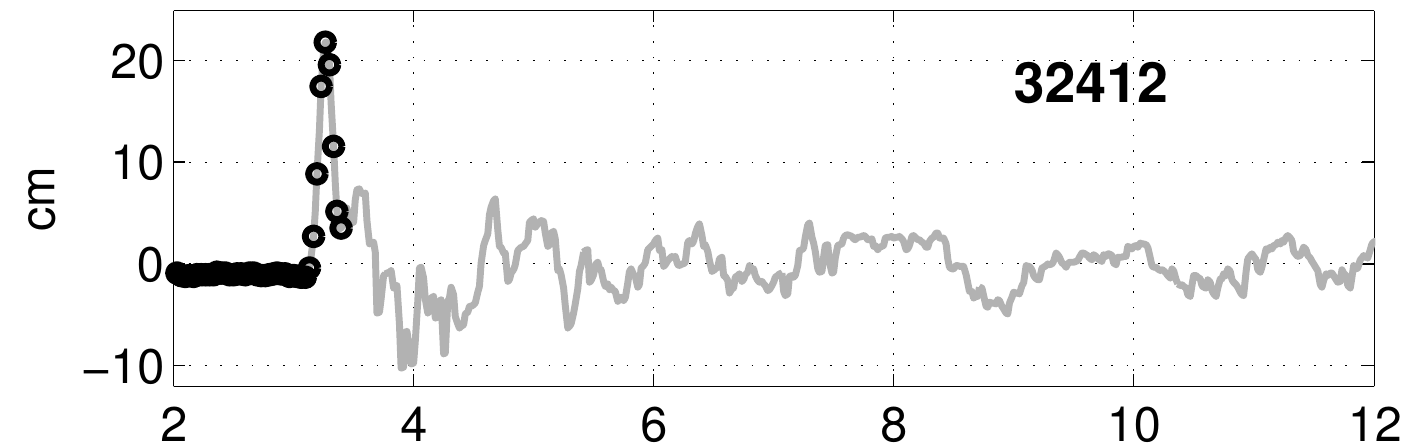} \\
	\includegraphics[width=\textwidth]{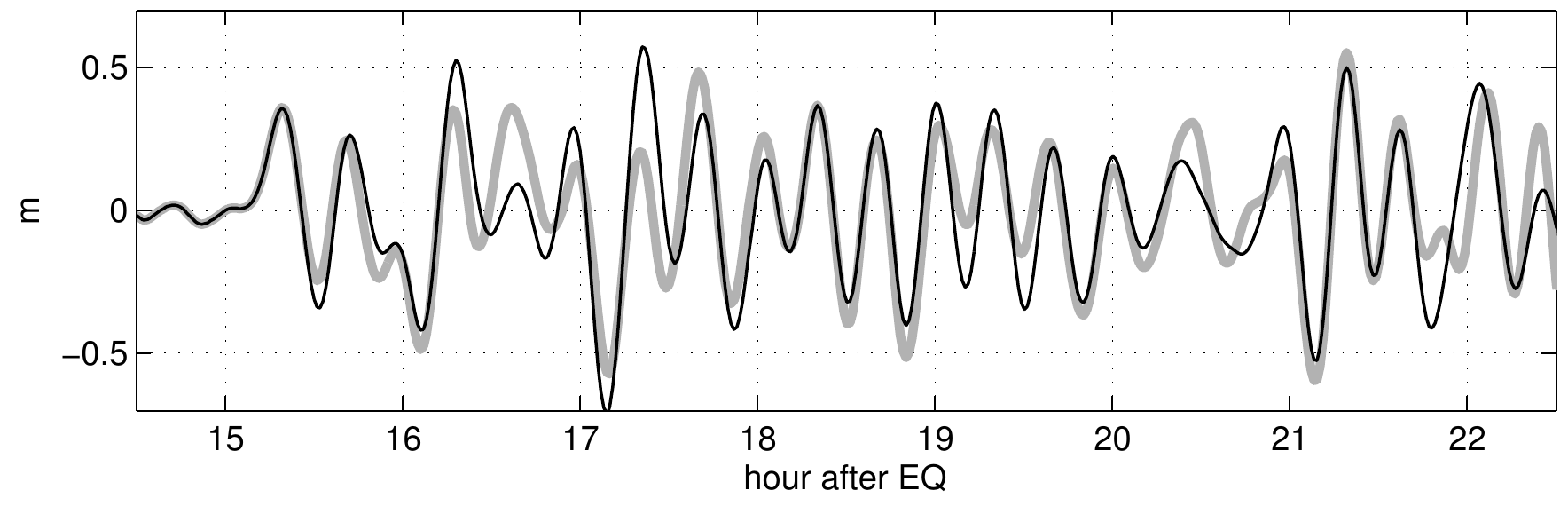} 
\end{tabular}
	\caption{ Top: record of the 2010 Chile tsunami at station 32412 (gray solid), and its measurements at 2-min intervals within the first 20-min of the tsunami recording at the station (black circles). Bottom: the tsunami prediction at Crescent City made with the long record at 32412 (gray thick), and with the early-available data (limited to the first 20 min) only (black thin).}
	\label{CC2}
	\end{center}
\end{figure} 

During a tsunami event, forecast would be attempted as soon as, say, first tsunami peak has passed the detector. Figure \ref{CC2} shows a prediction of the 2010 Chilean tsunami in Crescent City made with the first 20-min of the tsunami's recording at DART 32412, which contain the first wave peak. For comparison, the figure also shows a prediction generated with the longer time-history at 32412. The two predictions coincide for the first 20 min - the duration of the observations at 32412 shared by the two predictions.
Past the duration of the observations at the DART, without contribution from the later-arriving negative wave, the early, 20-min-based prediction slightly over-estimated the observed signal in the next hour, though still appears fairly accurate.
So the response-based forecast can be made early in the event, and then refined as more data become available.

Forecast timing is different between the PRF method and the inversion-based methods, since the two approaches feature different use of data.
The PRF approach takes a fraction of a second to make a forecast once DART observations are obtained, but the prediction is ``final" only for the duration of the available observations, especially early in the event.  
The prediction describes the site's response to the observed part of the wave train. 
Past the respective duration of the observations, the prediction lacks contribution from the waves which had not yet arrived at the detector(s). These later waves, if not small, can be included in the prediction later in the event.

For the inversion-based forecast, the observations need to accumulate until there is enough signal for the inversion  (say, 30 min - up to one full wave), followed by high-resolution simulations focusing on the locations of interest. 
Once this is done, the resulting prediction potentially describes the entire duration of the event.
However, since in the inversion process, the models are selected to fit only the first observed wave peak \citep{tang2012, perc}, the forecast accuracy might reduce past the duration of the site's response to the first wave(s) at the DART(s) being fitted. The inversion-based forecasts at all locations in Figures \ref{gages2011}-\ref{CC} produced with the SIFT system can be seen on NCTR's Events pages (\url{http://nctr.pmel.noaa.gov/honshu20110311/},  \url{http://nctr.pmel.noaa.gov/chile20100227/}). 

\section{Concluding remarks}
\label{remarks}

Predicting tsunami impacts at remote coasts largely relies on tsunami height measurements in an open ocean. 
In this work, we demonstrated that tsunami measurements can be used to formulate a boundary-value problem and numerically ``propagate" the wave directly from the detector(s), differently from the common practice of using the measurements to infer the tsunami source function. We suggested a formalism of specific response functions, which provide an instant solution to the above boundary problem, as long as non-linear effects are negligible. 
The tsunami predictions in the far field are generated as a response or a combination of responses to one or more tsunameters, with each response obtained as a convolution of real-time tsunameter measurements and a pre-computed Pulse Response Function. 

In the response formalism, wave field is assembled from wave beams assigned to individual detectors; and therefore the method's ability to resolve the wave azimuthal structure, or the directional diagram, depends on the detector's azimuthal coverage. Consequently, in the forecast examples presented, the method was applied to forecasting only the North America in the 2010 Chile tsunami, but almost the entire Pacific in the 2011-Tohoku event.
With the employment of the ocean depth correction method of Wang (2015), the tsunami arrival time errors common to tsunami models were largely eliminated in the computation of the response functions. 

For forecasting non-linear coastal environments, the response formalism can be used to generate offshore boundary inputs into fine-resolution grids focused on desired coastal sites. We also suggested an adaptation of the response formalism for providing direct forecasts in some non-linear environments such as a shallow harbor, although the limits of validity and the robustness of this adaptation are areas for further study.
Hindcasts of the two largest trans-Pacific tsunamis of the 21 century along the US West Coast and beyond have demonstrated high accuracy (not limited to the early arriving waves) of the PRF methodology.

As implied by the algorithm (section \ref{Nmath}) and/or the application examples (sections \ref{chile} and \ref{tohoku}), the proposed method has following benefits:
\begin{enumerate}
\item
The PRF method uses measurements for computing tsunami's forward propagation, which in mathematical terms represents a well-posed problem.
\item
The method  scales to any number of detectors, with the forecast accuracy and coverage gradually progressing as input from more detectors is incorporated.
\item
The method transmits the exact measurements for the entire duration of the event.
\item
The method has shown a promise for forecasting the later arriving waves.
\item
The method uses no assumptions about the EQ dynamics (such as an instantaneous co-seismic sea floor deformation), nor about the dynamics of the tsunami generation.
\item
Given an adequate detectors' coverage and pre-computed PRFs, the method can provide instant predictions in a tsunami event.
\end{enumerate}

At the same time, the PRF method might have more restrictive conditions of operation than the conventional methods have. The forecast coverage depends on the detector coverage -- in a seeming contrast to an inversion-based forecast with its potentially global coverage. In the latter case, however, the forecast accuracy depends on how well the source is constrained, that is, on the detector coverage as well. The PRF method relies on a pre-defined detector configuration.  DART stations rarely change their locations, but sometimes go out of service. Should a DART station become non-operational, PRF forecasts relying on this station cannot be accomplished. 
Moreover, with the present DART coverage, the applicability of the response formalism as developed herein is limited to tsunamis with compact sources, and to detectors receiving their wave trains from particular directions. The detectors therefore should be located outside the near field. This important limitation has not yet been quantified. However, this limitation has not been severe enough to preclude using this formalism with the major recent Pacific tsunamis. Future ``green repeaters" \cite[]{butler1} are likely to open more opportunities for application of the response formalism.

Thereby, depending on the detector coverage between the EQ location and the forecasted coast, the response formalism might provide a useful complement to existing forecasting technologies. While further work is needed to investigate in more detail different aspects of the response method, the existing tsunameter network already permits operating the response forecasting method with promising results, for some EQ locations and forecasted sites. 

\section{Acknowledgments}
We acknowledge NOAA/NDBC for providing the DART records, NOAA/NOS for providing the tide gauge records, Ocean Networks Canada for providing the NEPTUNE records, and NOAA/NCEI for providing the bathymetry used in the numerical simulations. Dmitry Nicolsky acknowledges support for his work from the state of Alaska.
\clearpage
\begin{appendix}

\section{Drafting the PRF Table}
\label{sPRFtbl}

\begin{SCfigure}
\centering
	\caption{EQ areas for PRF computation along a subduction zone east of Japan-Kurils-Kamchatka (dots) and local DART stations (triangles). }
		\includegraphics[width=0.6\textwidth]{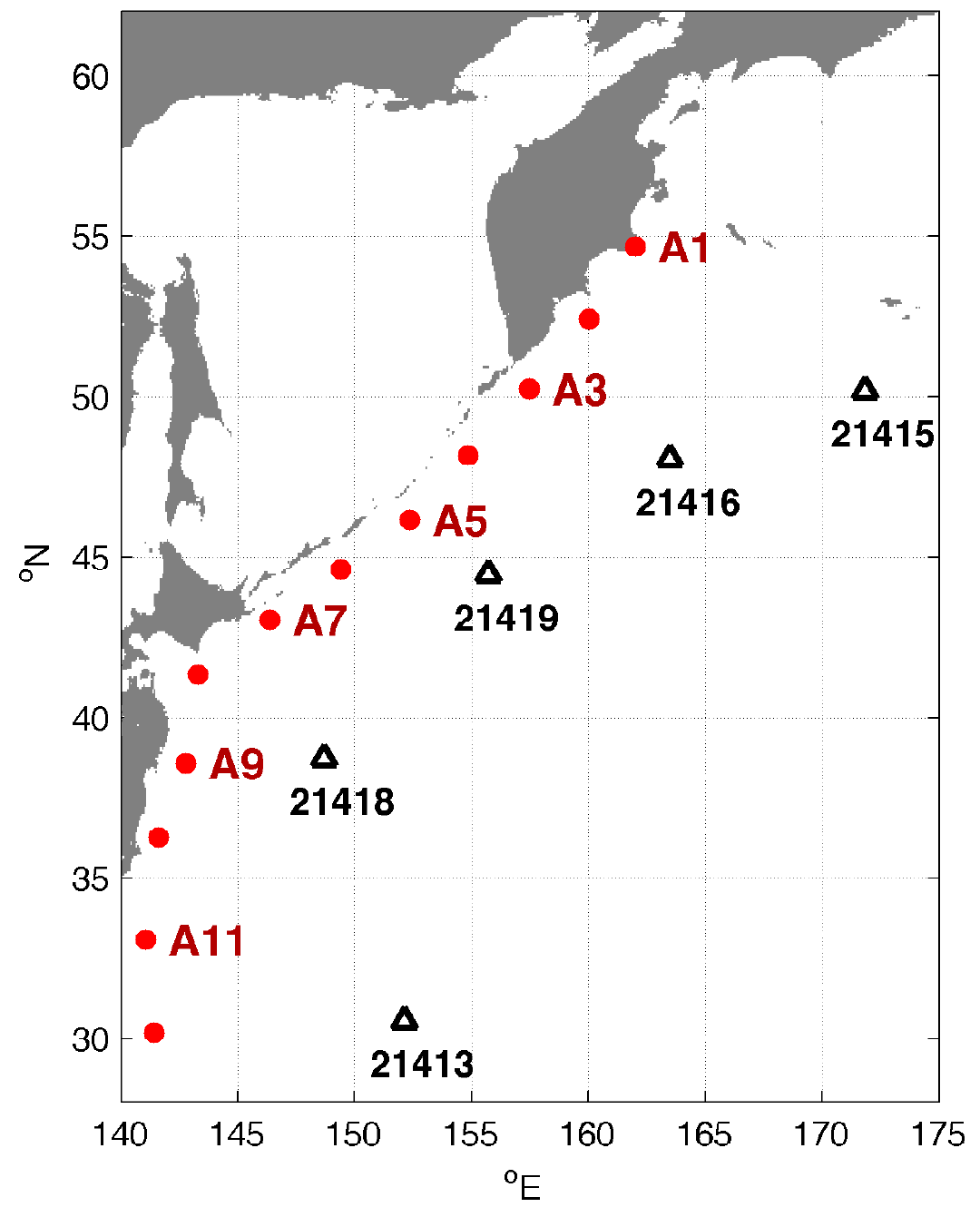}
	\label{kisz}
\end{SCfigure}
\begin{SCfigure}
\centering
	\caption{Structure of a PRF Table.}
		\includegraphics[width=0.5\textwidth]{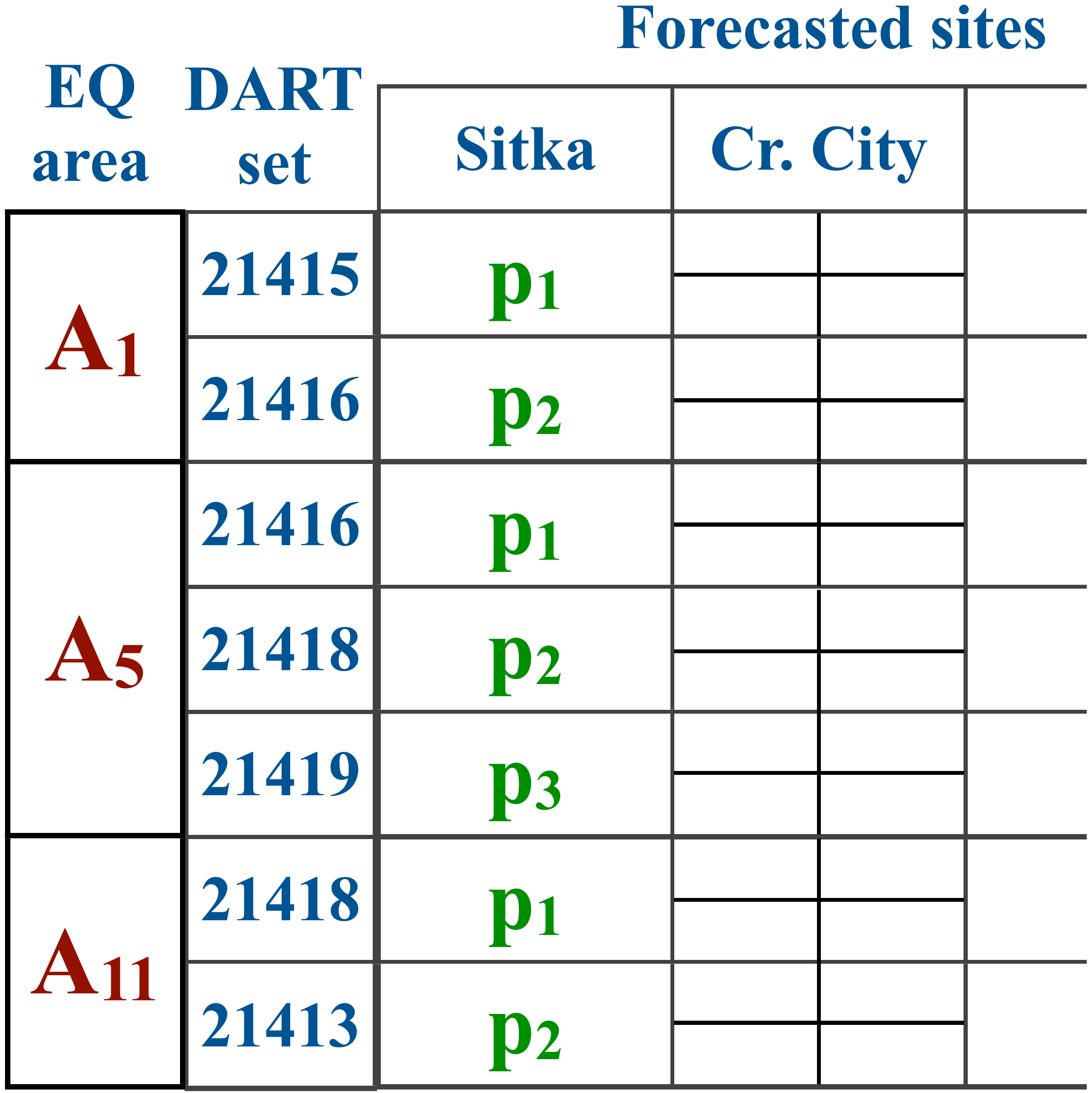}
	\label{PRFtable}
\end{SCfigure}

 
Practical implementation of the response method requires building tables of PRFs in a 3D parameter space: EQ location - tsunameter -  forecasted site. 
Here, we sketch such a table (a logical tree) for a hypothetical situation when a local emergency management wants to employ the PRF method as a complementary forecasting tool for evaluating tsunamis at selected coastal sites. They are especially concerned with tsunamis coming from a particular region, for instance, from an active tectonic plate interface east of Japan and Kurils. In section \ref{btest}, the same response set was successfully applied for forecasting tsunamis originating 300 km apart, as measured by the location of the EQ epicenters. So a 300 km distance is adopted as a length of the plate boundary served by the same responses. Dots in a map in Figure \ref{kisz} follow the plate interface with 300 km interval, and mark the EQ areas for PRF computations. The same responses will be used for tsunamis generated by EQs with epicenters in the same area, that is, within a 150-km distance (counted along the plate interface) from each dot. A corresponding PRF table in Figure \ref{PRFtable} lists these EQ areas in the first column. The second column lists DARTs assigned to different EQ areas. Other columns contain PRFs at forecasted sites. Given that the region is spanned by 12 EQ areas (Figure \ref{kisz}), with no more than 3 DARTs assigned to each area, then forecasting each coastal site requires no more than 36 PRF entries, with each PRF representing a time history, say, 20 h long after the pulse arrival, with 1-min sampling interval. With float-type (4 bytes per a sample) data, this amounts to under 170 KB per forecasted site, for the displayed tsunamigenic region. Forecasting sites with non-linear responses splits each PRF cell into several ones, containing scaled PRFs and levels $||P_a||$, as reflected in the table in a `Crescent City' column. 

Note, that the above PRF table is purely illustrative. Developing an actual table requires further study to delimit EQ areas, to find optimal DART sets to use with each EQ, and to research application of scaled PRFs at each location.  
\end{appendix}

\end{document}